\documentclass[aps,prb,showpacs,twocolumn,showkeys,amssymb,amsmath,nobibnotes,nofootinbib,superscriptaddress]{revtex4-1}
\usepackage[colorlinks=true,linkcolor=red]{hyperref}
\usepackage{amsmath}
\usepackage{amsfonts}
\usepackage{amssymb}
\usepackage{times}
\usepackage{color}
\usepackage{algorithm,algpseudocode}

\usepackage{graphicx}
\usepackage[sort&compress]{natbib}

\newcommand{\Tr}[1]{{\mathrm{Tr}}\left(#1\right)}

\newcommand{\Ord}[1]{{\mathcal O}\!\left(#1\right)}
\newcommand{\mean}[1]{\langle #1 \rangle}
\newcommand{\comm}[1]{}

\def\cP{{\mathcal P}}

\def\cV{{\mathcal V}}

\def\vv{{\bf v}}
\def\bx{{\bf x}}
\def\bz{{\bf z}}
\def\Min{{\mathrm{Min}}}

\begin{document}


\title{Random perfect lattices and the sphere packing problem}



\author{A.Andreanov}
\affiliation{Abdus Salam ICTP - 
Strada Costiera 11, 34151, Trieste, Italy}
\author{A.Scardicchio}
\affiliation{Abdus Salam ICTP - 
Strada Costiera 11, 34151, Trieste, Italy}
\affiliation{INFN, Sezione di Trieste -
via Valerio 2, 34127 Trieste, Italy}



\date{\today}

\begin{abstract}
Motivated by the search for best lattice sphere packings in Euclidean spaces of large dimensions we study randomly generated perfect lattices in moderately large dimensions (up to $d=19$ included). Perfect lattices are relevant in the solution of the problem of lattice sphere packing, because the best lattice packing is a perfect lattice and because they can be generated easily by an algorithm. Their number however grows super-exponentially with the dimension so to get an idea of their properties we propose to study a randomized version of the algorithm and to define a random ensemble with an effective temperature in a way reminiscent of a Montecarlo simulation. We therefore study the distribution of packing fractions and kissing numbers of these ensembles and show how as the temperature is decreased the best know packers are easily recovered. We find that, even at infinite temperature, the typical perfect lattices are considerably denser than known families (like $A_d$ and $D_d$) and we propose two hypotheses between which we cannot distinguish in this paper: one in which they improve Minkowsky's bound $\phi\sim 2^{-(0.84\pm 0.06) d}$,
and a competitor, in which their packing fraction decreases super-exponentially, namely $\phi\sim d^{-a d}$ but with a very small coefficient $a=0.06\pm0.04$. We also find properties of the random walk which are suggestive of a glassy system already for moderately small dimensions. We also analyze local structure of network of perfect lattices conjecturing that this is a scale-free network in all dimensions with constant scaling exponent $2.6\pm0.1$. 
\end{abstract}

\pacs{}

\maketitle

\section{Introduction}

Sphere packing is a classic problem with many connections to pure and applied mathematics (number theory and geometry~\cite{conway1999sphere}), communication theory~\cite{shannon1948bell} and physics~\cite{torquato2005random}. The statement of the problem is very simple: given an Euclidian space of dimension $d$ what is the densest spatial arrangement of impenetrable spheres? In a more formal way one seeks to find a maximum over all packings $$\phi_\text{best}(d)=\max_{\mathcal{P}\in S}\phi(\mathcal{P}).$$ Here $\mathcal{P}$ is a packing of spheres (an allowed configuration of the impenetrable spheres), $S$ is the set of all packings, and $\phi(\mathcal{P})$ is the fraction of space covered by the packing $\mathcal{P}$.

As is often the case with problems related to number theory, the simplest questions do not have simple answers. Despite over $200$ years of research the problem has only been solved for $d=2$~\cite{toth1940} and $d=3$~\cite{hales2005proof} (the famous Kepler's conjecture). The latter case has only been proven about fifteen years ago and required substantial amount of computer work. Although good and very good candidates for the best packings have been identified in higher dimensions (namely $\lesssim30$) our knowledge deteriorates quickly as dimensions become really high, say of order $~10^3$ where the problem becomes of interest to communications theory.

One the greatest challenges in the sphere packing problem is that no universal behavior is identifiable. Every dimension seems to be peculiar, with some dimensions being very special, like $8,12,24$. In the generic case there is no restriction on packings: they can be of any nature, ordered (crystalline breaking of translational symmetry) or even disordered. For relatively low dimensions, $d\leq 9$ the best (known) packings are all lattice packings, that is packings where spheres are placed at the vertices of a certain Bravais lattice (one particle per unit cell of the lattice). In $d=10$ for the first time, the best known packing is generated by a non-Bravais lattice~\cite{conway1999sphere}. Some recent works\cite{torquato2006new, zachary2011high} conjecture that in high enough dimensions completely disordered packings might win over regular ones.

To understand the degree of difficulty of the problem it is sufficient to mention that even finding good upper bounds  on best packing fractions uniformly valid for all dimensions resisted to all attacks so far. The one-hundred year old lower bound by Minkowsky only received linear improvements until today and an exponential improvement~\cite{torquato2006new, scardicchio2008estimates} only exists subject to an interesting but very strong conjecture\footnote{For recent considerations of the applications of statistical mechanics to Roger's bound\cite{rogers1958packing} see the work of Parisi\cite{parisi2008most}; see also~\cite{jin2010application,parisi2006amorphous,parisi2010mean}.}. Even worse, Minkowsky's bound is non-constructive, and no methods are known which would allow to construct a lattice which satisfies at least that bound in very high dimensions. Arguably the most important recent contribution in this respect has been given by the works \cite{cohn2003new, cohn2004optimality} in which the problem is reduced, for any given dimension, to an infinite linear programming problem. The technique is powerful --in 8 and 24 dimensions the bounds are saturated by the best known packing, proving hence their global optimality-- but has not yield an understanding of the problem for generic $d$.

Given the complexity of the generic case it might prove useful to consider a simpler version of the problem. One of them is the so called lattice packing problem, which restricts allowed packings to Bravais lattice packing only.~\footnote{Two slightly different terminologies are being used in mathematics and physics with respect to lattices: mathematicians differentiate between lattices and periodic sets, while physicists talk about Bravais and non-Bravais lattices.} Although the set of possible packings is severily reduced, exact results are only established up to $d\leq 8$, with $d=9$ case hopefully, closed in 2012.

In theory the lattice sphere packing problem is simpler, because it admits an explicit algorithmic solution~\cite{voronoi1908quelques} where one has to check a finite number of special lattices to find the best one. The best packing, in fact, is both a perfect and eutactic lattice (we give the characterization of these lattices later) and both the number of perfect~\cite{voronoi1908quelques,schurmann2009computational} and that of eutactic lattices is finite~\cite{ash1977eutactic} (hence the intersection is). This algorithm has been applied to dimensions $d\leq 8$ to systematically find all such lattices~\cite{lagrange1773recherches,gauss1840unter,korkin1877sur,barnes1957complete,jaquet1993enumeration,sikiric2007classification,riener2006extreme}. In this paper we will run a randomized version of the algorithm in dimensions 8 to 19 to generate large (up to several millions) set of perfect lattices in each dimensions and then study the statistical properties thereof. We will introduce a fictitious temperature to explore non-typical regions of the space of perfect lattices and get the best known packings.


\section{Lattices, perfect lattices and eutactic lattices}
\label{sec:voronoi}

\subsection{Notation}
In this paper we will consider only lattices or in Physics terminology Bravais lattices, namely lattices which have only one particle per unit cell. A generalization of our results to arbitrary but finite number of particles per unit cell will be discussed at the end of the paper. In our definitions and logic of discussion we will follow closely Schurmann~\cite{schurmann2009computational} although we will not pretend to achieve the same level of rigor.

We will define a lattice $A$, one particle per unit cell, in $\mathbb{R}^d$ by means the square matrix of the components of the $d$, $d$-dimensional linearly independent (basis) real vectors $\mathbf{e}^i$
\begin{equation}
A=\left(\begin{array}{cccc}
e^1_1 & e^1_2 & ... & e^1_d \\
e^2_1 & e^2_2 & ... & e^1_d \\
\vdots & \vdots & \ddots & \vdots \\
e^d_1 & e^d_2 & ... & e^d_d
\end{array}
\right).
\end{equation}
The points in the lattice are elements of the set
\begin{equation}
\Lambda=\{\mathbf{x}:\mathbf{x}=A \mathbf{z},\quad \mathbf{z}\in\mathbb{Z}^d/ \{\mathbf{0}\}\}.
\end{equation}
The associated symmetric, positive definite $d$-by-$d$ quadratic form $Q$ is defined by matrix multiplication as 
\begin{equation}
Q=A^T A.
\end{equation}
We will refer without difference to the quadratic form $Q$ or to the basis matrix $A$ when we talk about a lattice. The distance of a point $A\bz$ in the lattice is (here $T$ stands for transpose, both of a vector and of a matrix)
\begin{equation}
l=||\bx ||=\sqrt{\bz^T\ A^TA\ \bz}=\sqrt{\bz^TQ\bz},
\end{equation}
where $\bz^T Q\bz=\sum_{i,j=1}^d z_iQ_{ij}z_j$.

The notion of \emph{shortest vector} of a lattice is fundamental in the theory of lattices and allows one to connect to the theory of sphere packing. Namely define the \emph{arithmetic minimum} of a lattice $Q$ as square of the minimum length of a vector in the lattice
\begin{equation}
\lambda(Q)=\min_{\mathbf{z}\in\mathbb{Z}^d/ \{\mathbf{0}\}}\mathbf{z}^TQ\mathbf{z},
\end{equation}
and the set
\begin{equation}
\Min(Q)=\left\{\bz\in\mathbb{Z}^d:\ \bz^T Q\bz=\lambda(Q)\right\}.
\end{equation}
Let us point out that the set $\Min(Q)$ should contain at least two vectors (as $\bx$ and $-\bx$ have the same length) but for the ``interesting" lattices the cardinality of the set (known as the \emph{kissing number}) is usually much larger, sometimes even exponential in $d$. The maximum cardinality of $\Min(Q)$ over the set of $d$-dimensional lattices is an open problem in most $d$ and has been dubbed the \emph{kissing number problem}~\cite{conway1999sphere}.

The connection with the sphere packing problem is easily made. The largest non-overlapping spheres we can fit in a  lattice must have as radius half the length of the shortest vectors of $Q$. Considering that the volume of a unit cell is $\det A=\sqrt{\det Q}$ we have the maximum fraction of space covered by a sphere packing $Q$ is the ratio of the volume of this sphere divided by the volume of the unit cell:
\begin{equation}
\phi(Q)=B_d \frac{(\sqrt{\lambda(Q)}/2)^d}{\det(Q)^{1/2}}.
\end{equation}
where $B_d$ is the volume of a $d$-dimensional unit sphere
\begin{equation}
B_d=\frac{2 \pi^{d/2}}{d\ \Gamma(d/2)}.
\end{equation}
A strictly related quantity is the Hermite constant of $Q$ (in terms of which the packing fraction can be expressed)
\begin{equation}
H(Q)=\frac{\lambda(Q)}{\det^{1/d}(Q)}
\end{equation}
In the following we will also use another indicator that we will call ``energy" as a target function to minimize with the introduction of a temperature:
\begin{equation}
e(Q)=-\frac{1}{d}\log(\phi(Q)).
\label{def:energy}
\end{equation}
Minkowksy's bound ensures that this quantity is bounded on the best lattices even in the limit $d\to\infty$.

The \emph{lattice sphere packing problem} (henceforth LSP problem) in $d$ dimensions is the problem of finding the maximum of $\phi(Q)$ (or $H(Q)$) among all the $d$-dimensional lattices. The problem is solved for $d=1,...,8$~\cite{lagrange1773recherches,gauss1840unter,korkin1877sur,barnes1957complete,jaquet1993enumeration,sikiric2007classification,riener2006extreme} and $d=24$~\cite{cohn2003new,cohn2004optimality} \emph{only}. 

\subsection{Perfect lattices}

We will now concentrate on a subset of lattices which turns out to be fundamental in the solution of the lattice sphere packing problem: the \emph{perfect lattices}.

A lattice is named \emph{perfect} iff the projectors built with its shortest vectors span the space of symmetric $d$-by-$d$ matrices. So for a perfect lattice $Q$ let $Z$ be the cardinality of $\Min(Q)$ and let $\vv_a\in\Min(Q)$, $a=1,...,Z$ ($Z$ is also called the \emph{kissing number} of a lattice). Let $M$ be any symmetric $d$-by-$d$ matrix there exist a set of reals $\mu_a$ such that:
\begin{equation}
M=\sum_{a=1}^Z\mu_a\vv_a \vv_a^T.
\end{equation}
For example take the square lattice in $d=2$:
\begin{equation}
Q_{sq}=\left(\begin{array}{cc}1 & 0 \\0 & 1\end{array}\right)
\end{equation}
the shortest vectors are
\begin{equation}
\Min(Q_{sq})=\{ (1,0),(0,1)\}
\end{equation}
and the projectors are 
\begin{equation}
P_1=\left(\begin{array}{cc}1 & 0 \\0 & 0\end{array}\right),\quad P_2=\left(\begin{array}{cc}0 & 0 \\0 & 1\end{array}\right),
\end{equation}
which do not span the space of symmetric matrices. Therefore the square lattice is \emph{not} a perfect lattice.

Instead, consider the hexagonal lattice
\begin{equation}
\label{eq:qhexlat}
Q_{hex}=\left(\begin{array}{cc}2 & 1 \\1 & 2\end{array}\right).
\end{equation}
It has three shortest vectors (of length\footnote{We remind the reader that the length of a vector is $(\bx^T Q\bx)^{1/2}$.} $\sqrt{2}$)
\begin{equation}
\Min(Q_{hex})=\{(1,0),(0,1),(1,-1)\}
\end{equation}
and the corresponding projectors are
\begin{equation}
P_1=\left(\begin{array}{cc}1 & 0 \\0 & 0\end{array}\right),\quad P_2=\left(\begin{array}{cc}0 & 0 \\0 & 1\end{array}\right),\quad P_3=\left(\begin{array}{cc}1 & -1 \\-1 & 1\end{array}\right),
\end{equation}
and the reader can verify that they form a basis for symmetric 2-by-2 matrices (one can easily form linear combinations of the $P$'s to obtain the identity and two of the three Pauli matrices). Note that the number of shortest vectors of a perfect lattice is bounded from below by $(d+1)d$ since this is twice the smallest possible number of projectors that can span the space of symmetric matrices (the dimension of space of symmetric matrices). So in the previous example we could have said beforehand that the square lattice is not perfect but we should have checked anyway that the hexagonal lattice was indeed perfect.\footnote{In d=2 it turns out that 6 (3 shortest vectors and their opposite $-\bx$) is also the maximum \emph{kissing number} achievable among lattices (and among general point patterns too).}

\comm{To grasp an intuition on the difference between a perfect lattice and a non-perfect one, we perturb the square lattice by a small symmetric matrix:
\begin{equation}
\label{eq:qsqe}
Q_{sq+\epsilon}=\left(\begin{array}{cc}1 & \epsilon \\ \epsilon & 1\end{array}\right).
\end{equation}
For sufficiently small $\epsilon$ the shortest vectors are unchanged:
\begin{equation}
\Min(Q_{sq.+\epsilon})=\{(1,0),(0,1)\}
\end{equation}
but the determinant does change and therefore the Hermite constant is
\begin{equation}
H(Q_{sq+\epsilon})=\frac{1}{(1-\epsilon^2)^{1/2}}\simeq 1+\frac{1}{2}\epsilon^2.
\end{equation}
The change of the packing fraction is of order $\epsilon^2$. For the hexagonal lattice instead the change of packing fraction is of $\Ord{\epsilon}$ as can be seen using numerical tools (the derivative of the Hermite constant turns out to be $0.27216...$). So \emph{an order $\epsilon$ modification of a perfect lattice leads to a $\Ord{\epsilon}$ modification of the packing fraction.} Perfect lattices do not have ``flat directions", so to speak.}

Voronoi proved~\cite{voronoi1908quelques,schurmann2009computational} that perfect forms are vertices of the Ryshkov polyhedron\footnote{the Ryshkov polyhedron is not a finite polyhedron but it is a \emph{locally finite} polyhedron. This difference turns out to be immaterial here.} defined as a set of forms $Q$ whose shortest vector is larger than a given value:
\begin{equation}
\cP_{\lambda}=\{Q:\lambda(Q)\geq\lambda\}
\end{equation}
where the actual value of $\lambda$ (as far as $\lambda>0$) is immaterial as the axis can be rescaled freely. Therefore we can reduce the sphere packing problem on $\cP_\lambda$, hence constraining to forms with $\lambda(Q)=\lambda$ without any loss by finding 
\begin{equation}
H=\frac{\lambda}{\inf_{Q\in\cP_\lambda}\det^{1/d}(Q)}.
\end{equation}
The number of vertices of the Ryshkov polyhedron, and hence of perfect forms is (up to isometries that we define below) finite (a small subset of all the lattices in any given dimension $d$).

\begin{figure}[htbp]
\begin{center}
\includegraphics[width=0.45\columnwidth]{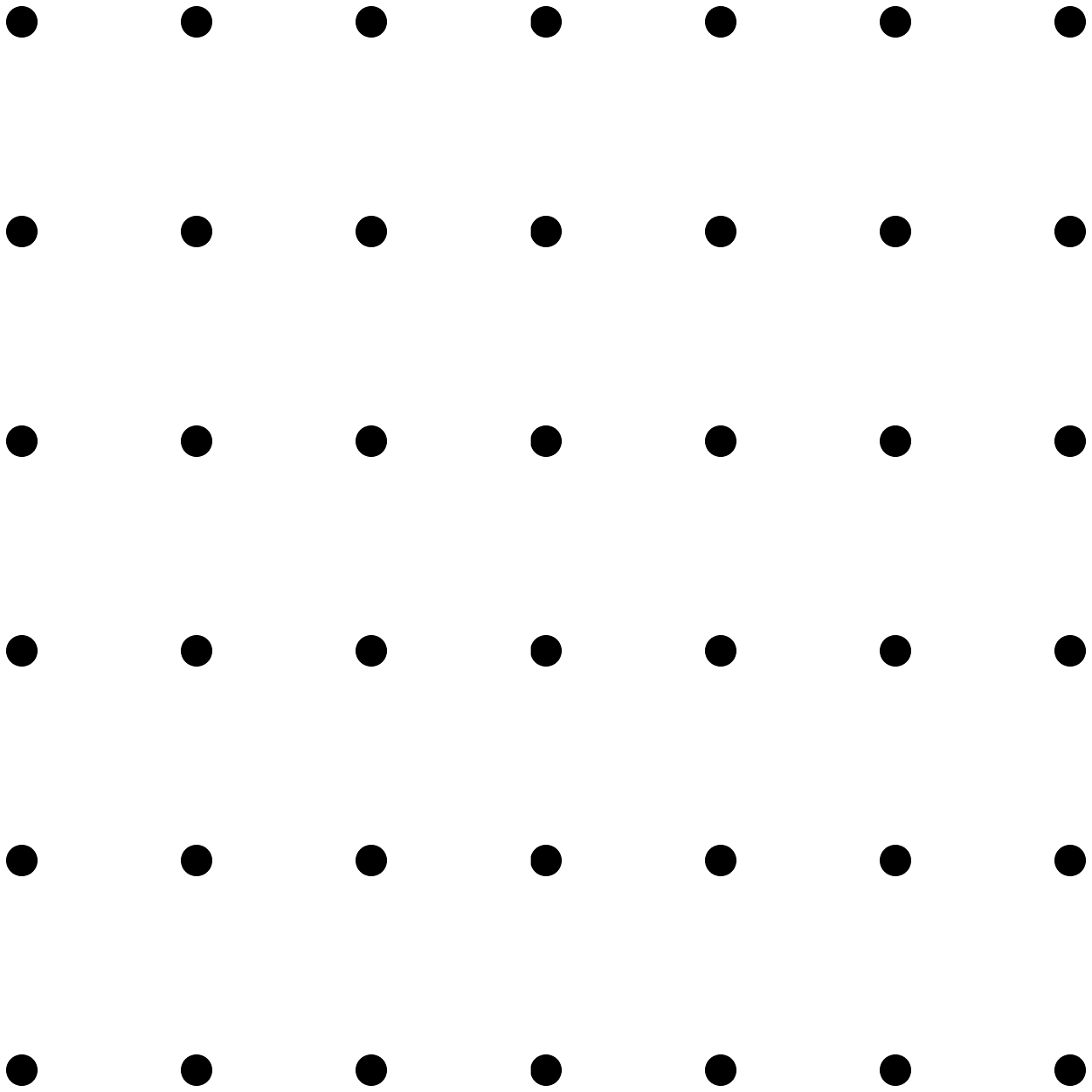}
\hspace{0.5cm}\includegraphics[width=0.45\columnwidth]{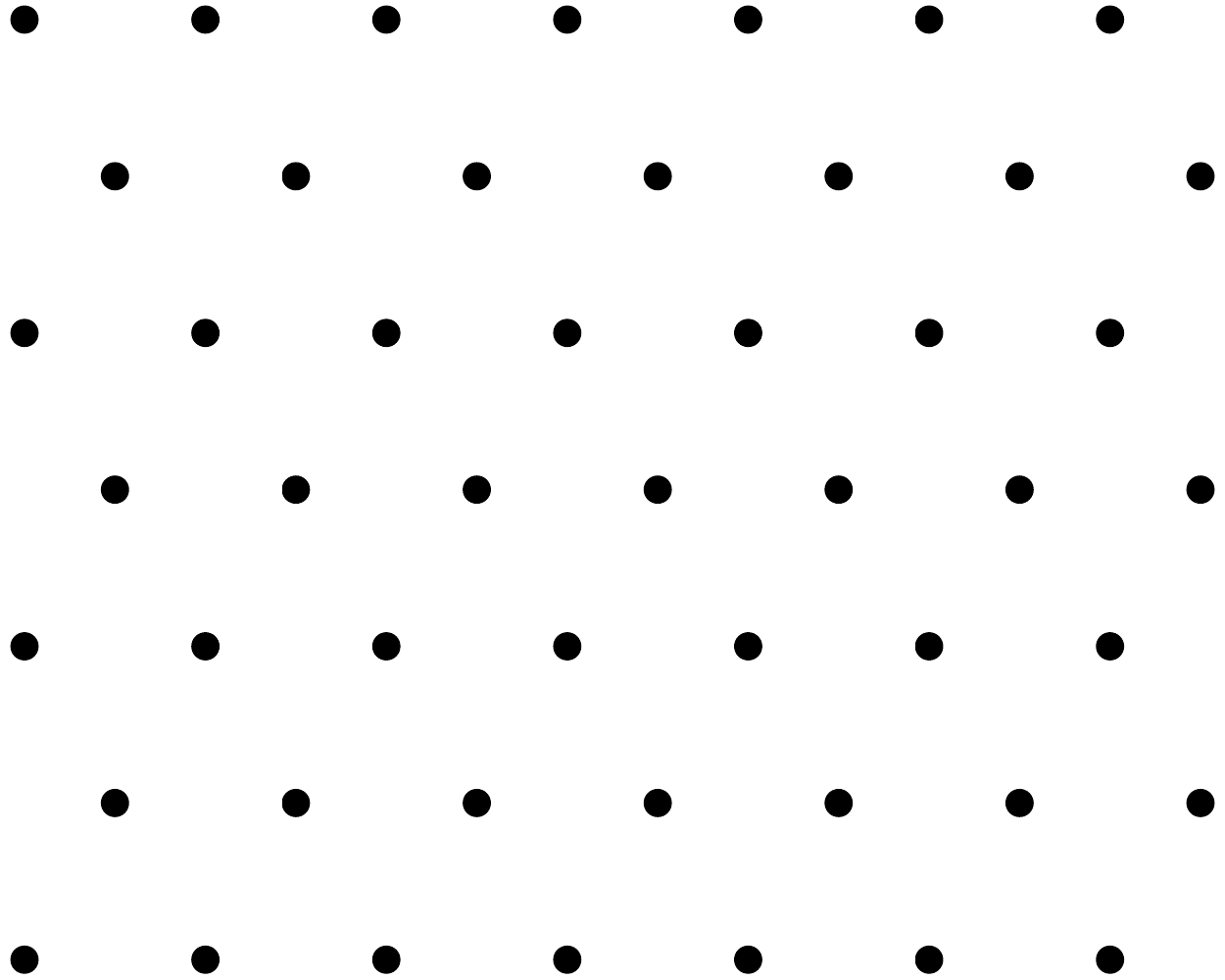}
\caption{\emph{Left. } Square lattice. \emph{Right.} Hexagonal lattice (also known as triangular lattice).}
\end{center}
\end{figure}

The main result which gives importance to perfect lattices in the context of the LSP problem is the classic \emph{Voronoi's theorem} which can be stated as follows:

\vspace{0.5cm}
{\bf Theorem:} \emph{ the best lattice sphere packing is a perfect lattice.}
\vspace{0.5cm}

The proof (which we do not give here, see\cite{schurmann2009computational}) follows if one shows that \emph{$\det^{1/d}(Q)$ does not have stationary points inside the Ryshkov polyhedron}. This in fact implies that the minimum of $\det(Q)$ and the maximum of $\phi$ (or $H$) occur on the vertices of the polyhedron, hence on perfect lattices.

Therefore the problem of LSP is reduced to finding all the perfect lattices and comparing their packing fractions: it becomes a problem for a computer to solve.\footnote{In principle LSP is an algorithmically solvable problem even without restricting to perfect lattices, since the number of Bravais lattices is finite in any dimension (for example there are 14 such lattices in 3 dimensions, 64 in 4 dimensions and the number should rapidly increase with $d$). However the mere enumeration of Bravais lattice is an unaccomplished task in $d\geq 7$ and to our knowledge no algorithm for generating them sequentially exists. Restricting the problem to perfect lattices simplifies it considerably.} Unfortunately (or maybe, fortunately) things are not so easy as they might seem. Indeed the number of perfect lattices grows very fast with the dimension (probably faster than exponential, as we will argue later) and the task of finding them all has been completed up to $d=8$ (where they are 10916). For $d=9$ has found $5\cdot 10^5$ forms~\cite{schurmann2012perf9} but the conjectured total number should be around $2\cdot 10^6$.

\subsection{Isometry of lattices}

A lattice admits many equivalent representations in terms of quadratic forms $Q$: one can rotate the lattice or replace its basis vectors with their independent linear combinations. This equivalence is captured by notion of \emph{isometry}:

\vspace{0.5cm}
{\bf Definition:} \emph{Lattices $Q$ and $Q^\prime$ are isometric if there exists a matrix $U\in\text{GL}_d(\mathbb{Z})$ and $c\in\mathbb{R}$ such that $$Q^\prime = c\,U^t\,Q\,U.$$}
\vspace{0.5cm}

\noindent Another name in use is \emph{arithmetical equivalence}. For example the hexagonal lattice $Q_{hex}$ given by Eq.~\eqref{eq:qhexlat} has an equivalent representation $$Q_{hex}^\prime=\left(\begin{array}{cc}2 & -1 \\-1 & 2\end{array}\right)$$ which is isometric to $Q_{hex}$ with isometry matrix $$U=\left(\begin{array}{cc}1 & -1 \\-1 & 0\end{array}\right).$$

A practical way of checking if a given pair of forms are isometric was developed in~\cite{plesken1997computing} where one uses backtrack search to construct an isometry matrix (if this exists). However most of the times it is sufficient to check if some criteria (like the number of shortest vectors) are satisfied before running the generic code which can be quite slow in high dimensions.

\subsection{Eutaxy}
The last concept that we need for our investigation is that of \emph{eutactic} lattice. This is not strictly necessary for understanding our results in this paper but it gives a suggestive connection with the theory of spin glasses which we plan to investigate as a continuation of this work. Eutactic lattices cannot be improved (as we will prove below) by an infinitesimal transformation of the matrix base and therefore are local maxima of the packing fraction. Their number also grows with the dimension $d$ and one is then led to think that in high enough dimensions this phenomenon is reminiscent of the landscape of a mean-field spin glass free energy \cite{mezard1987spin}.

Given a perfect form $Q$ we can always write (since it is a symmetric, nonsingular matrix) its inverse $Q^{-1}$ in terms of the projectors built on its shortest vectors
\begin{equation}
\label{eutaxy-def}
Q^{-1}=\sum_{x\in\Min(Q)}\alpha_x\ \bx\bx^T
\end{equation}
(here $\bx\bx^T$ is the matrix with elements $x_i x_j$).

{\bf Definition:} A \emph{eutactic} form is one for which one can choose all the above $\alpha_x>0$. An equivalent definition is that $Q^{-1}$ is in the interior of the Voronoi domain of the perfect form $Q$, defined as
\begin{equation}
\cV(Q)={\mathrm{cone}} \{ \bx \bx^T:\ \bx\in\Min(Q)\},
\end{equation}
the cone in the space of forms generated by the projectors built with the shortest vectors of $Q$.

The Hermite constant (or packing fraction) of an eutactic form can only be decreased by any infinitesimal change of the form. In fact, by using the identity 
\begin{equation}
\Tr{(\nabla\det Q)A}=\det(Q)\Tr{Q^{-1}A}
\label{eq:derdet}
\end{equation}
we obtain, to first order in $\delta Q=Q'-Q$ where $Q'\in\cP_{\lambda(Q)}$ (so the length of the minimal vectors is unchanged):
\begin{equation}
H(Q+\delta Q)=H(Q)-\frac{\lambda/d}{\det^{1/d}(Q)}\Tr{Q^{-1}\delta Q}< H(Q)
\end{equation}
where the inequality follows from:
\begin{equation}
\Tr{Q^{-1},Q'-Q}=\sum_{\bx\in\Min(Q)}\alpha_{\bx} (\bx^TQ'\bx  -\bx^TQ\bx) > 0,
\end{equation}
as $Q'\in\cP_{\lambda(Q)}$ and $\alpha_{\bx}>0$.

It follows then that \emph{a perfect and eutactic lattice is a local maximum of} $H$ from which
\vspace{0.5cm}

{\bf Theorem:} \emph{perfect and eutactic (PE) lattices are local maxima of the Hermite constant and hence of the packing fraction} 

and therefore

{\bf Corollary:} \emph{the best packing lattice is both perfect and eutactic.}

\vspace{0.5cm}

The concept of eutaxy is extended to arbitrary lattices with introduction of \emph{weakly-eutactic}, \emph{semi-eutactic} and \emph{strongly-eutactic} lattices. Weakly-eutactic lattices satisfy Eq.~\eqref{eutaxy-def} with real coefficients $\alpha_x$, semi-eutactic lattices have $\alpha_x\geq0$ (i.e.\ some of the coefficients in Eq.~\eqref{eutaxy-def} are zero) and finally strongly-eutactic lattices are eutactic lattices with all $\alpha_x$ equal. Recall that by definition a perfect lattice is (at least) weakly-eutactic since $\mathbf{x}\mathbf{x}^T$ span the space. The interest in strongly-eutactic lattices comes from the fact they are also the best packers locally among lattices with arbitrary number of particles per unit cell~\cite{schurmann2010perfect}.

The problem of determining eutaxy class of a form admits an efficient solution: given a form, its eutaxy class - non-eutactic, weakly-eutactic, semi-eutactic or  strongly-eutactic - can be decided by solving a sequence of linear programs~\cite{riener2006extreme} and therefore is of polynomial complexity with respect to the number of shortest vectors (which, however can grow as fast as an exponential of $d$).

Summarizing, the take home messages of this section are that the maximum of the packing fraction over lattices in any given dimension is attained by one of the PE lattices, of which there is a finite number (in any given $d$) and that each of the PE lattices is a local maximum. This characterization is extremely powerful but still does not prevent us from having to find all perfect lattices and checking which ones are eutactic and which are not. There is a simple and efficient way to generate perfect lattices but there is not (as far as we know) a similarly efficient way to generate eutactic~\cite{batut2001classification,bergé1996classification} or PE lattices. One should first generate perfect lattices and then check them for eutaxy. The simple and efficient way to generate perfect lattices is given by Voronoi algorithm, which we review in the following section.

\section{Voronoi's algorithm and its randomization}


We have now reduced the problem of finding the best lattice packing to that of finding the best lattice packing among perfect and eutactic lattices. We need a way to generate all the perfect lattices, select the eutactic ones and look at the most dense among them.
The first task is accomplished by the {\bf Voronoi algorithm}~\cite{voronoi1908quelques,schurmann2009computational,martinet2003perfect} that we now describe. 

\emph{Start} with a perfect form $Q$.
\begin{enumerate}
\item Find all the shortest vectors $\bx\in\Min(Q)$, and the inequalities describing the cone $\cP(Q)$
\begin{equation}
\cP(Q)=\{Q'|\ \forall \bx\in\Min(Q): \bx^TQ'\bx\geq 0\}
\end{equation}
\item Find all the extreme rays of the polyhedral cone $\cP(Q)$. Call them $R_1,...,R_k$.
\item Create the forms $Q_i=Q+\alpha_i R_i$, choosing rational numbers $\alpha_i$ such that the new form $Q_i$ is again perfect.
\item Check for isometries and repeat from Start with each of the genuinely new $Q_i$'s.
\end{enumerate}

In this way we are guaranteed to find all the perfect forms. If we check for isometry with previously found forms the algorithm will at a certain point terminate, its output being the list of all perfect forms in a given dimension. The extreme rays of an $n$-dimensional polyhedral cone are the half-lines at which at least $n-1$ inequalities are binding ($n=d(d+1)/2$ here). The bottleneck of the algorithm is finding all the extreme rays $R_i$ of a given lattice $Q$~\cite{bremner2009polyhedral} (or more rigorously of the Voronoi domain $\cV(Q)$), which, since the number of minimal vectors can be quite large (as much as exponential in $d$) can be a complicated linear programming problem. The generic version of this problem is known as a \emph{polyhedral representation conversion} problem in polyhedral computation community and its complexity is currently unknown~\cite{avis2009polyhedral,bremner2009polyhedral}. All the forms generated from a given form $Q$ are called \emph{neighbors of} $Q$ and the graph consisting of perfect forms linked to their neighbors is called the \emph{Voronoi graph} of perfect forms in a given dimension $d$. Importantly, the graph is connected and starting from any vertex one can at least in principle reach any other vertex of the graph~\cite{voronoi1908quelques,schurmann2009computational,martinet2003perfect}.

\begin{figure}[htbp]
\begin{center}
\includegraphics[width=0.45\columnwidth]{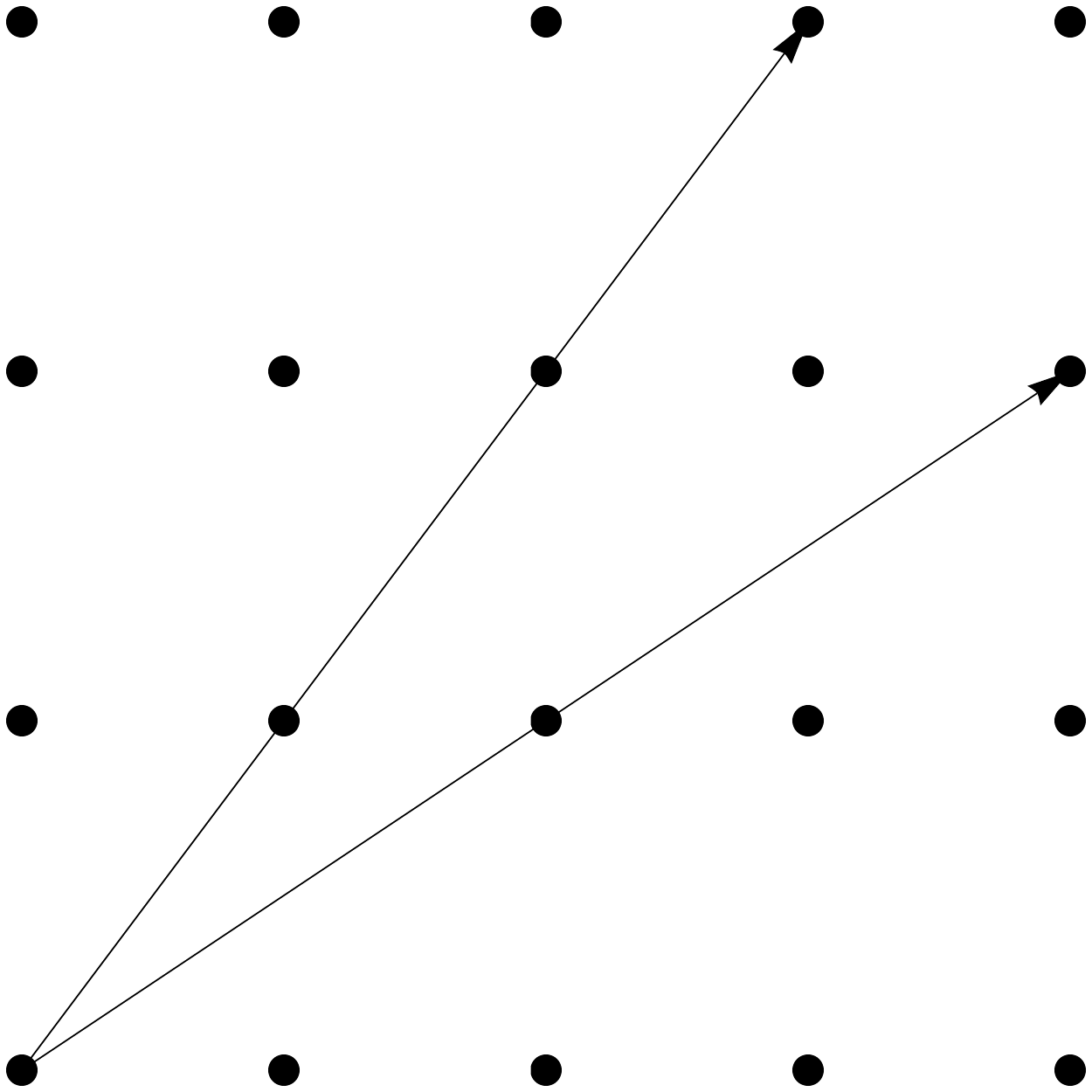}
\includegraphics[width=0.45\columnwidth]{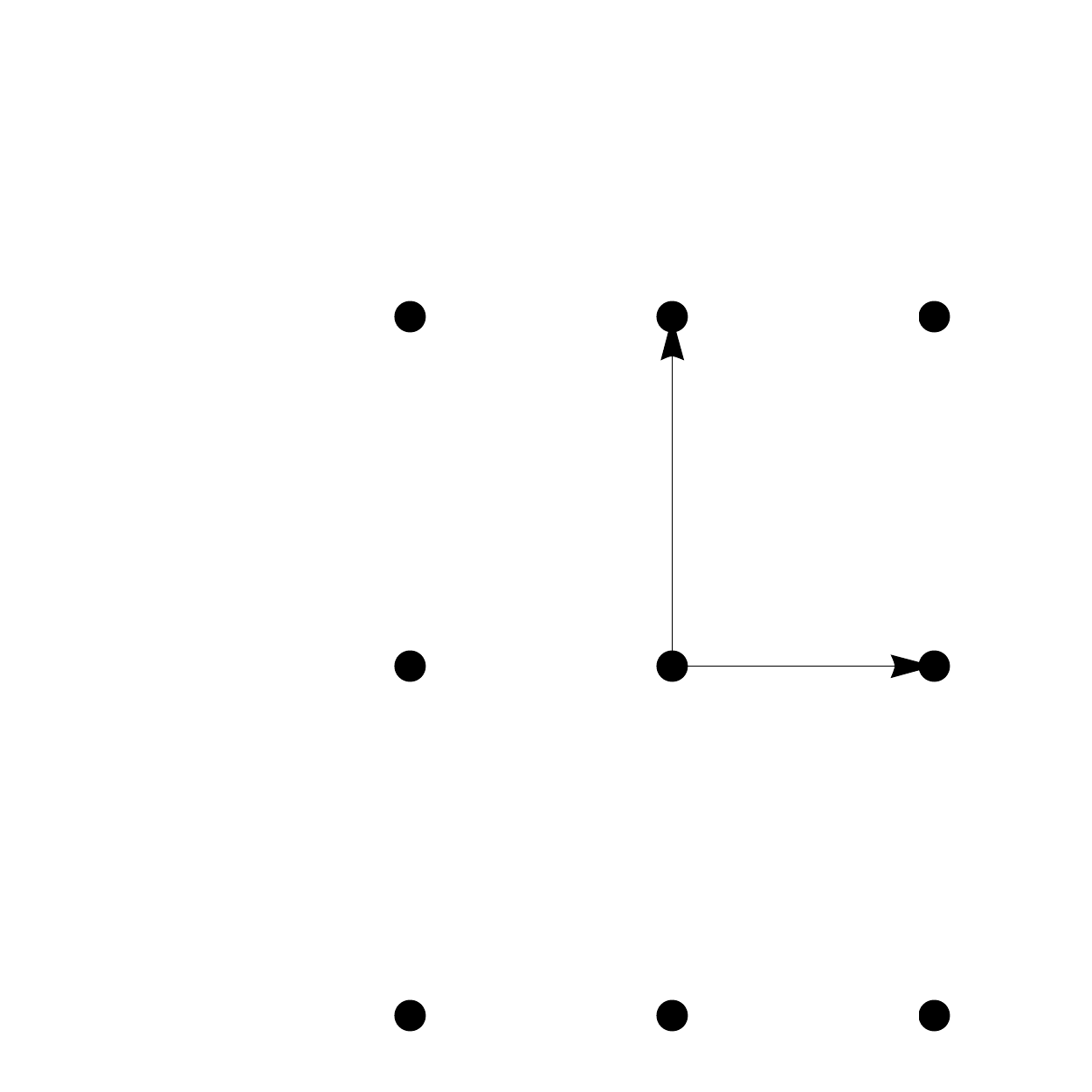}
\caption{Example of lattice reduction for a square lattice: random initial basis (\emph{left}) where basis vectors have large norms. After lattice reduction (\emph{right}) one gets "short" basis vectors.}
\label{fig:lll-square-lattice}
\end{center}
\end{figure}

Thus generated lattices might (and often do) have generating forms with rather large norms of basis vectors. For example while we know there is just a single perfect form in $d=2$. However a plain random walk would generate forms with entries growing as a function of step number. To remedy this problem we use the fact that for a given lattice its basis can be transformed to an equivalent basis but with reduced basis vector norms. Figure~\ref{fig:lll-square-lattice} illustrates this idea for square lattice. The exact transformation which reduces the norms to the smallest possible value is expensive and we use a inexact one known as LLL-reduction after the names of the authors~\cite{lenstra1982factoring} to produce equivalent representations of lattices with rather short basis vectors. Technically we apply the LLL-reduction on every newly generated form: this extra step allows us to generate forms with relatively small entries. Coming back to $d=2$ case we find just $3$ distinct forms (all of which are isometric). It is worth pointing out that the probability of generating isometric forms becomes much less relevant for higher dimensions and completely irrelevant for $d\geq 13$. The LLL reduction is also a subset of isometry testing and actually removes the most trivial isometries.

In order to focus on higher dimensions we propose to \emph{randomize} Voronoi's algorithm, namely to introduce a randomized subroutine to find an extreme ray $R_i$. In this way we do not have to find all the extreme rays but just pick one and move in that direction.

We do the following: we slice the cone with a plane, in this way the extreme rays become vertices of a polyhedron. We then define a random linear cost function
\begin{equation}
\label{eq:randomLP}
f(Q')=\sum_{i,j=1}^{d}A_{ij}Q'_{ij}
\end{equation}
where the $A_{ij}$ are gaussian random variables and we solve the corresponding linear programming problem $\max_{Q'\in \cP(Q)} f(Q')$. Linear functions are necessarily maximized at the vertices of the polyhedral region and therefore in this way we select randomly an extreme ray, which gives a neighbor of $Q$. The gaussian distribution of the $A_{ij}$ induces a distribution on the frequency each neighbor is visited which is far from uniform (a vertex is visited more often if, in the polyhedron it is surrounded by facets with relatively large surface). We will discuss later our attempts to make more uniform this distribution.

We have now defined the random generation of a new neighbor of $Q$ so in order to define a random walk we need to define the rules 
for accepting or rejecting said moves.

\section{Monte-Carlo procedure and the Voronoi graph}

It is clear that if we are only interested in the structure of the Voronoi graph we should run a random walk as unbiased as we can. Of course the most naturally unbiased algorithm would ideally generate any neighbor with equal probability. However this would be equivalent to finding all the neighbors for every perfect lattice; this problem can be incredibly difficult and it has been solved only for $d\leq 8$~\cite{sikiric2007classification}, with a large use of computer resources, so we do not attempt to solve it here.


\subsection{A warm-up: simple cases $d\leq 7$}

As a warm up we study very low dimensions: for $d\leq 7$ the problem of enumeration of perfect lattices is relatively simple due to small number of non-isometric perfect forms $\mathcal{N}$:
\begin{center}
\begin{tabular}{ | c | c | c | c | c | c | c | c | }
\hline
Dimension & $1$ & $2$ & $3$ & $4$ & $5$ & 6 & $7$ \\
\hline
$\mathcal{N}$ & 1 & 1 & 1 & 2 & 3 & 7 & 33 \\
\hline
\end{tabular}
\end{center}
The problem is completely trivial for $d\leq 3$ since there is a single perfect lattice (up to isometries). For $d=4,5$ enumeration is trivial: our code finds the other forms on the first steps. Less trivial cases are $d=6$ and $d=7$ with $7$ and $33$ perfect forms respectively. It takes about one thousand steps to find all $7$ forms in $d=6$. In $d=7$ we recover $32$ forms after $10^6$ steps.
\begin{figure}
\begin{center}
\includegraphics[width=0.45\columnwidth]{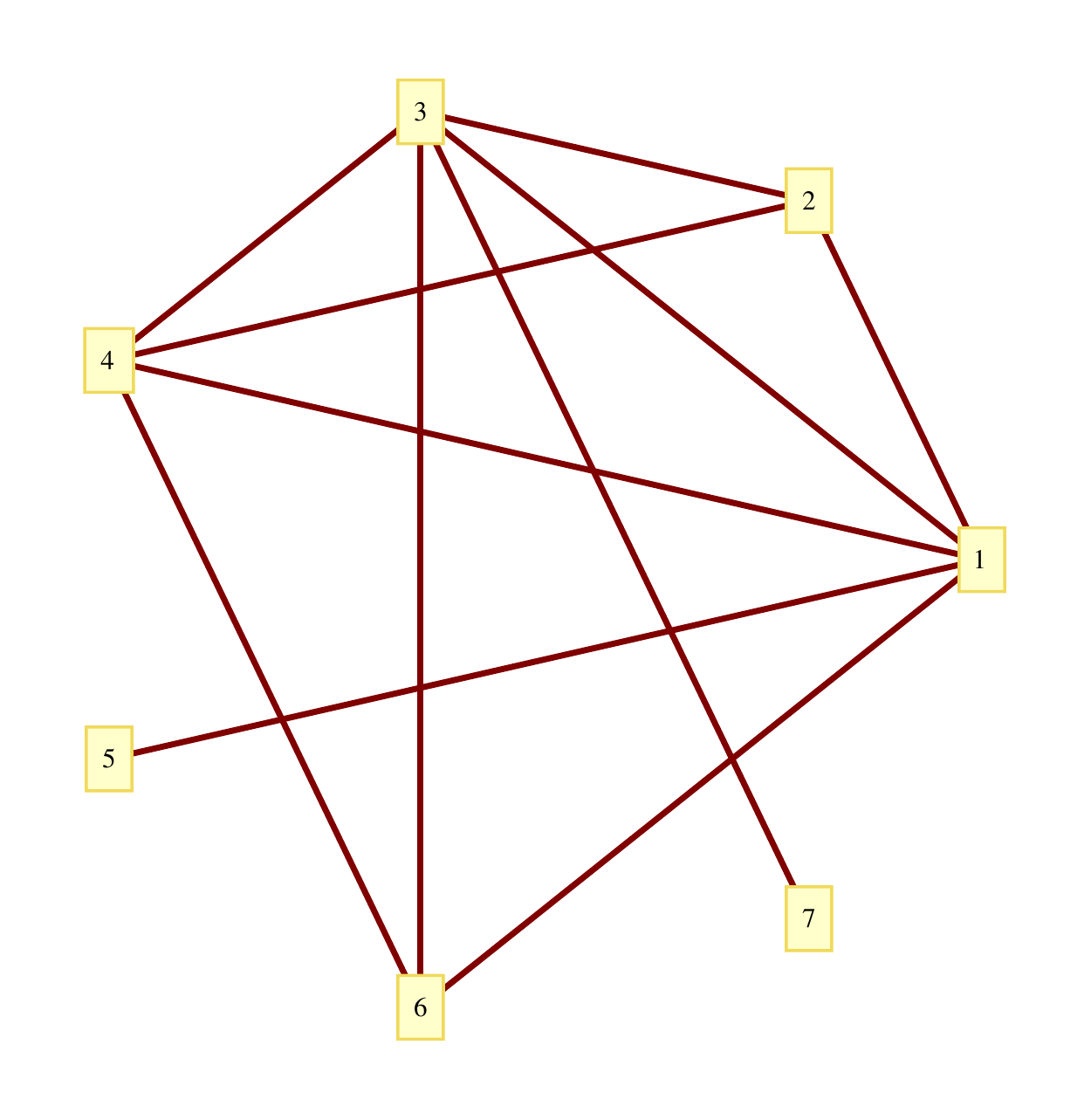}
\includegraphics[width=0.45\columnwidth]{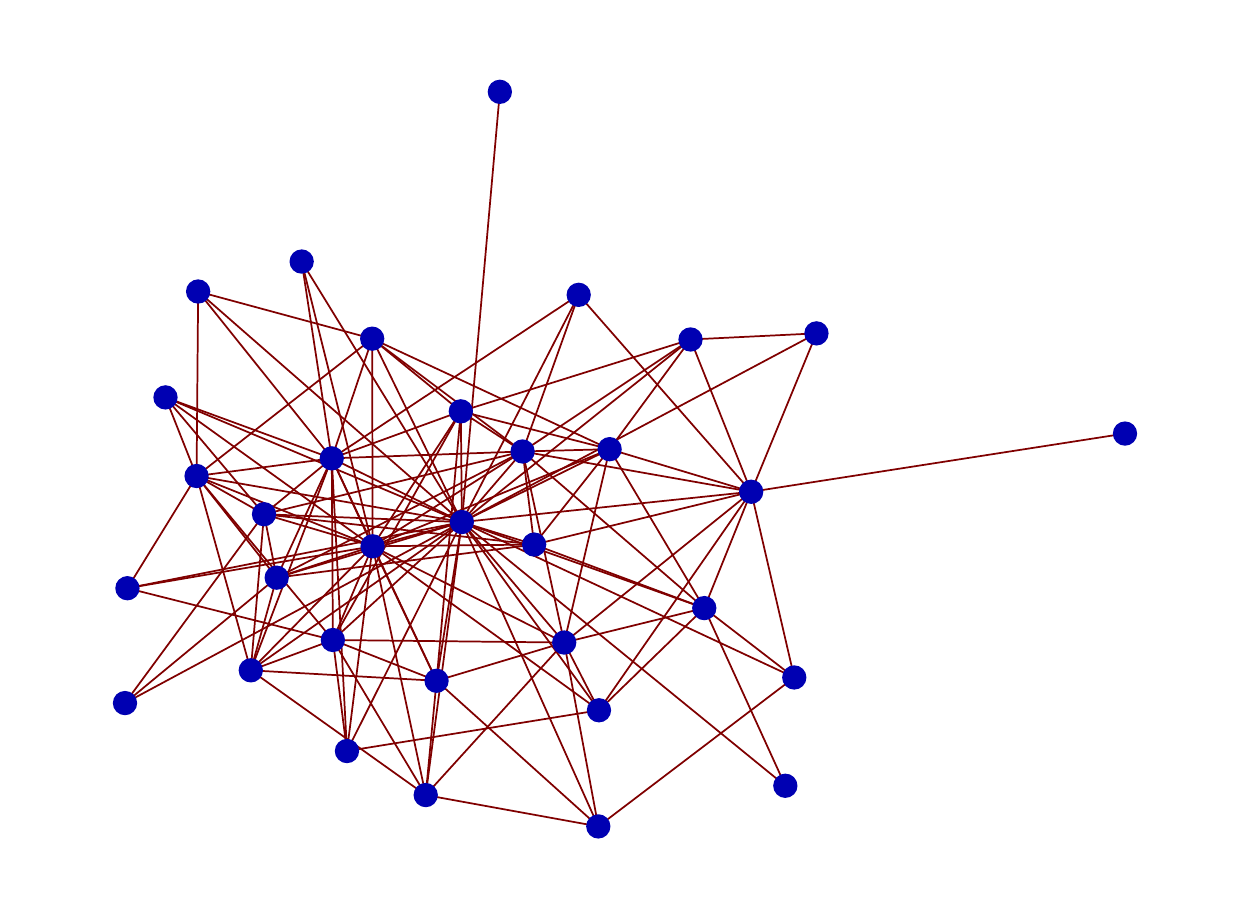}
\caption{\emph{Left} The Voronoi graph in $d=6$; vertex $1$ is $E_5$, vertex $3$ is $D_6$, vertex $7$ is $A_7$. \emph{Right} The Voronoi graph in $d=7$: there are just $33$ perfect forms. The central point is $E_7$: it is connected to all the other vertices but $A_7$, which is the rightmost vertex of the graph.}.
\label{fig:perf7graph}
\end{center}
\end{figure}

\subsection{Properties of the $d=8$ and $d=9$ Voronoi graphs}

We compare the random walk on the exact Voronoi graph as found in~\cite{sikiric2007classification} with the numerical results of the previously described randomized Voronoi algorithm.

The Voronoi graph for $d=8$ is quite an interesting object if seen through the lens of statistical mechanics of random graphs. We unveil here only a small set of observations. The number of vertices is the number of perfect forms, namely 10916, and we put an edge whenever two forms are Voronoi neighbors. The most connected form is the densest packing $E_8$, which has 10913 neighbors, and it is interesting to notice that the distribution of the connectivity of the graph follows quite closely a power law decay (a so-called \emph{scale-free} network) for $c\lesssim 2\ 10^3$. Over this 3-orders of magnitude range we can fit the connectivity distribution by the law
\begin{equation}
\label{eq:connd8}
p(c)\propto c^{-(2.7\pm 0.1)},
\end{equation}
which defines a critical exponent. We will see that this is also the case in $d=9$.
\begin{figure}[htbp]
\begin{center}
\includegraphics[width=0.9\columnwidth]{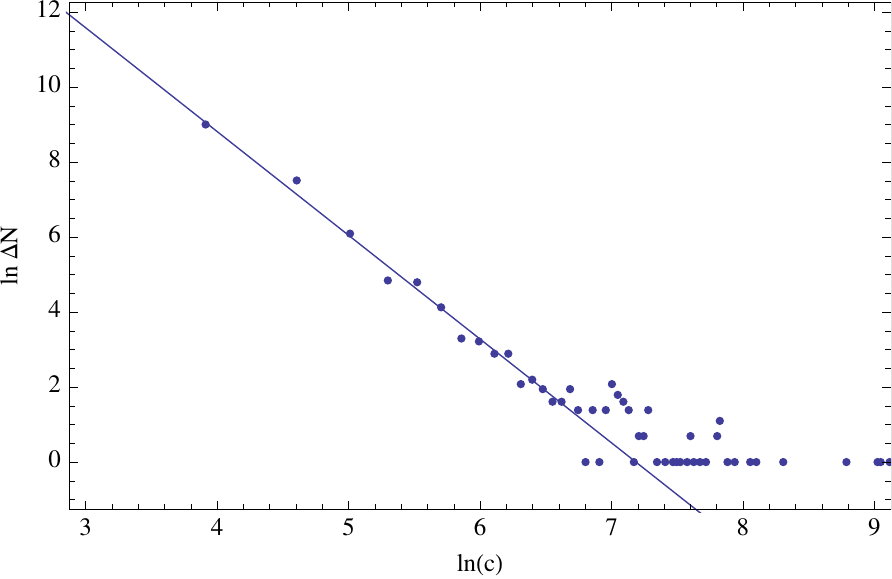}
\includegraphics[width=0.9\columnwidth]{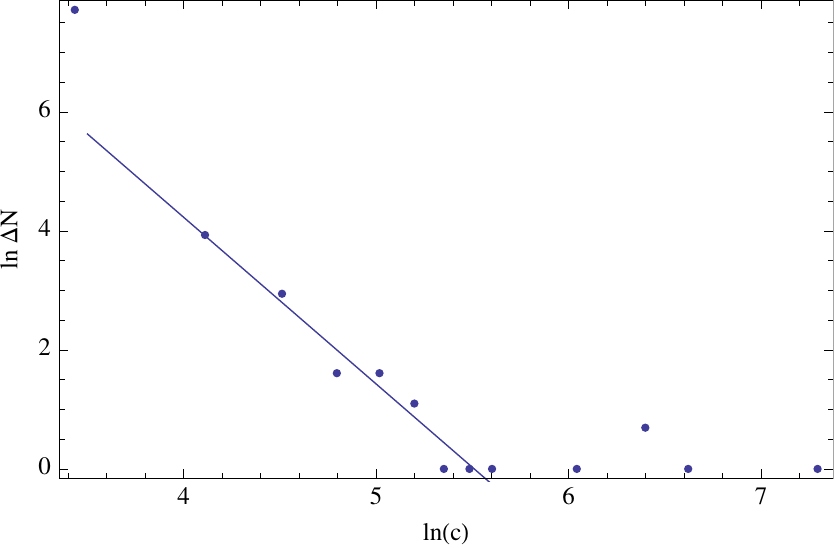}
\caption{\emph{Top:} The distribution of the connectivity of the $d=8$ Voronoi graph, exact results and \emph{Bottom:} the same distribution sampled with the randomized Voronoi algorithm. $\Delta N(c)$ is the number of perfect lattices which have connectivity between $c-50$ and $c$. The power-law fit is described in equation (\ref{eq:connd8}). In general, an underestimation by the random walk of the connectivity of the nodes is observed but a power law fit still works well, and the power law is compatible with the exact result (see text).}
\label{fig:connd8}
\end{center}
\end{figure}

It follows from the large connectivity of $E_8$ that an unbiased random walk on this graph would visit $E_8$ a large number of times. By running a completely unbiased random walk on the \emph{exact} Voronoi graph in 8 dimensions we find that $E_8$ should be visited about $1.6\%$ of the times (this has to be compared with an average of $1/10916\simeq 0.01\%$). In our algorithm we see however that this number is much larger: $E_8$ is visited around $80\%$ of the time. This means that our algorithm is biased towards lattices with higher connectivity even more than an unbiased random walk is. This has to do with the large surface occupied by facets of the Ryshkov polyhedron enclosed by rays generating $E_8$.

This is a common feature in any dimension: the densest lattices are reached quite fast by our randomized algorithm even in absence of any a priori bias towards them. The balance between the increase in the attractivity of the best packers and the increase in the size of the graph allows one to stumble upon the densest lattice up to $d=12$ with a few hundred trials without having to bias the random walk towards the densest lattices. Moreover, as a typical scale-free network, the diameter of the Voronoi graphs will be quite small, scaling as the logarithm of the number of vertices divided by the logarithm of the average connectivity.

We now discuss the results of our randomized algorithm in $d=8$. We find, as said, that $80\%$ of the times is spent on $E_8$. The remaining $20\%$ of the time is divided among the remaining lattices. Every time a lattice is visited an isometry test is run against the previously visited lattices. If it is new, it is added to the list; in any case a link between the two lattices is added to the list of edges in the graph. In this way, in $10^6$ runs we generate around $3\cdot 10^3$ non-isometric perfect lattices (out of 10916). This might be taken as a measure of the importance of isometry as well as of the dominance of $E_8$ in 8 dimensions.

In $d=9$ we run the randomized Voronoi algorithm for $10^6$ steps and we generate around $6\cdot 10^4$ non-isometric perfect forms. We recall that in $d=9$ the Voronoi graph is conjectured to be made of around $2\cdot 10^6$ inequivalent perfect forms. We hence find in this case that the importance of isometry is much reduced. We will see that in higher dimensions the isometry test becomes irrelevant as randomly generated forms turn out to be almost always non-isometric.

\begin{figure}[htbp]
\begin{center}
\includegraphics[width=0.9\columnwidth]{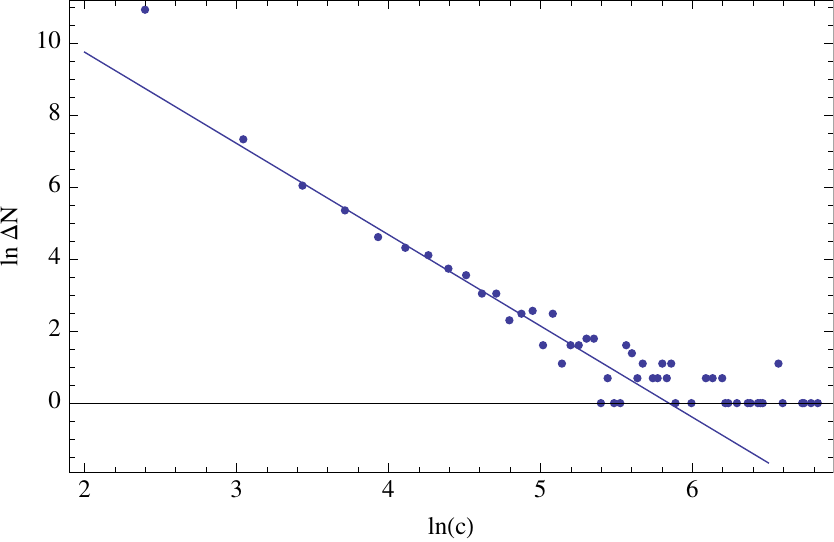}
\caption{The distribution of the connectivity of the $d=9$ Voronoi graph estimated by the random walk. $\Delta N(c)$ is the number of perfect lattices which have connectivity between $c-50$ and $c$. The power-law fit is described in equation (\ref{eq:connd9})}.
\label{fig:connd9}
\end{center}
\end{figure}

By looking at the distribution of the local connectivity we see that also in this case a power-law distribution is the best fit over 3 orders of magnitude:
\begin{equation}
p(c)\propto c^{-2.5\pm 0.1}.
\label{eq:connd9}
\end{equation}

We also observe the same slight overestimate of the fraction of low connectivity graphs we saw in $d=8$. This is due (as in other dimensions) to the fact that the in order to assign a connectivity $c$ to a graph the random walk has to visit said graph at least $c$ times. There's no proved estimate of number of perfect lattices (size of the Voronoi graph) as a function of dimension. The sequence looks as $1,1,1,3,7,33,10916,\sim2\cdot10^6,\dots$  and suggests a superexponential growth, for example like $e^{A\,d^2}$. Consequently the number of steps required for an accurate estimation of connectivity grows rapidly. This means that for dimensions higher than 9 a different strategy has to be used.

However, after observing the similarity between the two exponents for the connectivity distribution and checking our random walk results against the exact results in $d=8$ it is nothing but tempting to \emph{conjecture} that the \emph{Voronoi graph is a scale-free random network in any dimension} and that \emph{the exponent of the distribution of the connectivity is around $2.6$.}

\begin{figure}[htbp]
\begin{center}
\includegraphics[width=7.5cm]{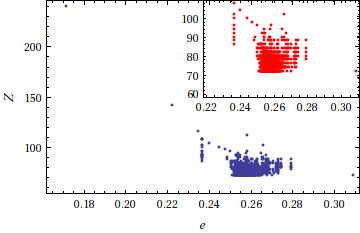}
\includegraphics[width=7.5cm]{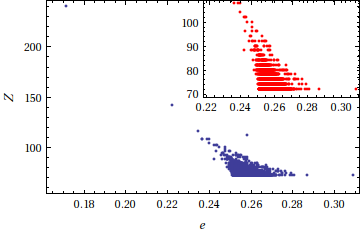}
\caption{\emph{Top} Kissing number \textit{vs} energy ($d=8$), generated set. \emph{Bottom} Kissing number \textit{vs} energy ($d=8$), exact data. The insets show same plots with kissing numbers $Z\leq110$. In both cases the best packer and kisser is alone in the upper left of the figures.}
\label{fig:ekiss8}
\end{center}
\end{figure}

\begin{figure}[htbp]
\begin{center}
\includegraphics[width=7cm]{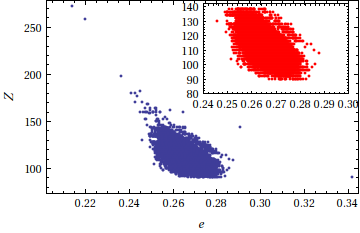}
\caption{Kissing number \textit{vs} energy ($d=9$). The inset shows detailed plot for kissing numbers $Z\leq140$.}.
\label{fig:ekiss9}
\end{center}
\end{figure}

One can also plot (see fig.\ \ref{fig:ekiss8} and \ref{fig:ekiss9}) the joint distribution of kissing number and energy observing how the best packers have largest kissing number and they are both rare events with respect to the typical distribution. This phenomenon is constant across all dimensions.

\section{Biasing the random walk with a temperature}
\label{sec:biasrwbeta}

Following a common trick in statistical mechanics we introduce a temperature $\beta$ as a Lagrange multiplier for the packing fraction. We therefore would like to define a statistical ensemble described by the partition function:
\begin{gather}
\label{eq:partfunc}
Z  = \sum_{Q}\mu(Q)\,e^{-\beta\,d^2\,e(Q)}\\
e(Q) = -\frac{1}{d}\log\phi(Q)\nonumber
\end{gather}
where $Q$ is a perfect lattice in $d$ dimensions and $\mu(Q)$ is the measure induced on the space of perfect lattices by the solution of the linear program (\ref{eq:randomLP}),\footnote{In practice we introduce the temperature on the random walk via Montecarlo sampling but since we cannot assure that the detailed balance holds for our randomized Voronoi algorithm we cannot ensure that we are quantitatively sampling the partition function above. For the purpose of this paper this is a minor point.} namely, $\mu(Q)$ is the fraction of times the lattice $Q$ is visited when the random walk described in the previous section is run. We also defined \emph{energy of a packing} $e(Q)$ so in (\ref{def:energy}) that it is a quantity of order $1$ for the best packings which have packing fraction decreasing exponentially in dimension. Quite conveniently the best packings translate into packings with lowest energy i.e.\ ``ground states". The normalization for the temperature is due to the expectation that for the densest lattices $\log(\phi)\sim d$ (as both upper and lower bounds predicts) and we need the exponent to be order of the number of degrees of freedom, namely $\sim d^2$. 

By lowering the temperature we expect to explore the regions of the Voronoi graph in which lattices are denser.

\section{Results}

Below we present the numerical results generated by random walks described above and their interpretation.

\subsection{Aims}

The generation procedure is inherently stochastic and we do not aim at generating complete sets of perfect lattices in a given dimension. As we already mentioned we have discovered $32$ and approximately $3\cdot 10^3$ forms after $\sim10^6$ runs in $d=7$ and $d=8$ respectively. The number of discovered forms in $d=8$ increases with extra runs, although a complete enumeration would require a huge number of runs. 

Such a huge number of perfect lattices suggests a statistical approach so that properties of typical or even dense lattices can be extracted from a subset of the complete set. Thus our goal is rather to generate sufficiently large, representative sets of perfect forms in a given dimension which would allow us to understand typical properties of perfect lattices and spot any universal pattern behind.

The fact that we are dealing with relatively large sets of forms together with the stochastic nature of the generating procedure allows to introduce empirical distributions of various characteristics of lattices. We are going to focus mainly on two quantities: \emph{energy} which was defined above and \emph{kissing number}. Both quantities are of interest with respect to the best packings. We will analyze their statistical properties, in particular their distributions and moments on the ensemble generated by the random walk.

We have generated random walks (both simple and biased) in dimensions from $8$ to $19$. Complexity of computation gradually increases with dimension as does typical running time to generate sufficiently representative set of lattices. Runnning times vary from about an hour in $d=8,9$ to $5-7$ days in $d=19$ to generate $~5\cdot10^4$ lattices. Higher dimensions, i.e. $d\geq 20$ are accessible, the difficulties encountered being rather of technical than conceptual nature.

\subsection{Random walk at infinite temperature}

We have first performed runs in different dimensions at infinite temperature which correspond to plain random walks: departing from initial lattice one computes a random neighbour and hops there. It is natural to think that this way one generates typical perfect lattices~\footnote{Remember that there is already a bias builtin into generation of neighbours!}. The walk terminates after a finite number of steps $N$ have been made. The averages $\langle\dots\rangle$ are simple summations normalised by $N$.

Typically $A_d$ was used as a starting point of a random walk for $d\leq 12$ and $D_d$ was used for $d\lesssim 15-16$ since the energy of $A_d$ becomes too high. In even higher dimensions ($d>16$) the energy of $D_d$ itself becomes too  high for $D_d$ to be a good starting point and we used different initial lattices with better packing fractions which we generated by chain runs, that is first running a random walk starting at $D_d$ and then picking a suitably dense lattice as a starting point for a new random walk.

As already mentioned above our randomised code is biased towards denser lattices and it doesn't sample all lattices uniformly like a complete enumeration would do (this effect is \emph{on top} of the bias given by the larger connectivity of the densest lattices). It is instructive to compare our results to exact data. Unfortunately the latter are only known for $d<9$~\footnote{Enumeration in $d=9$ is in progress, see~\cite{schurmann2012perf9}. Partial results are avalaible, but due to nature of the enumeration procedure they are biased and cannot be directly compared to our data} and there are too few perfect lattices for our approach to be benefitial for $d<8$. So we start by comparing energy and kissing number distributions as sampled by our code and their exact values in $d=8$.
\begin{figure}[htbp]
\begin{center}
\includegraphics[width=0.9\columnwidth]{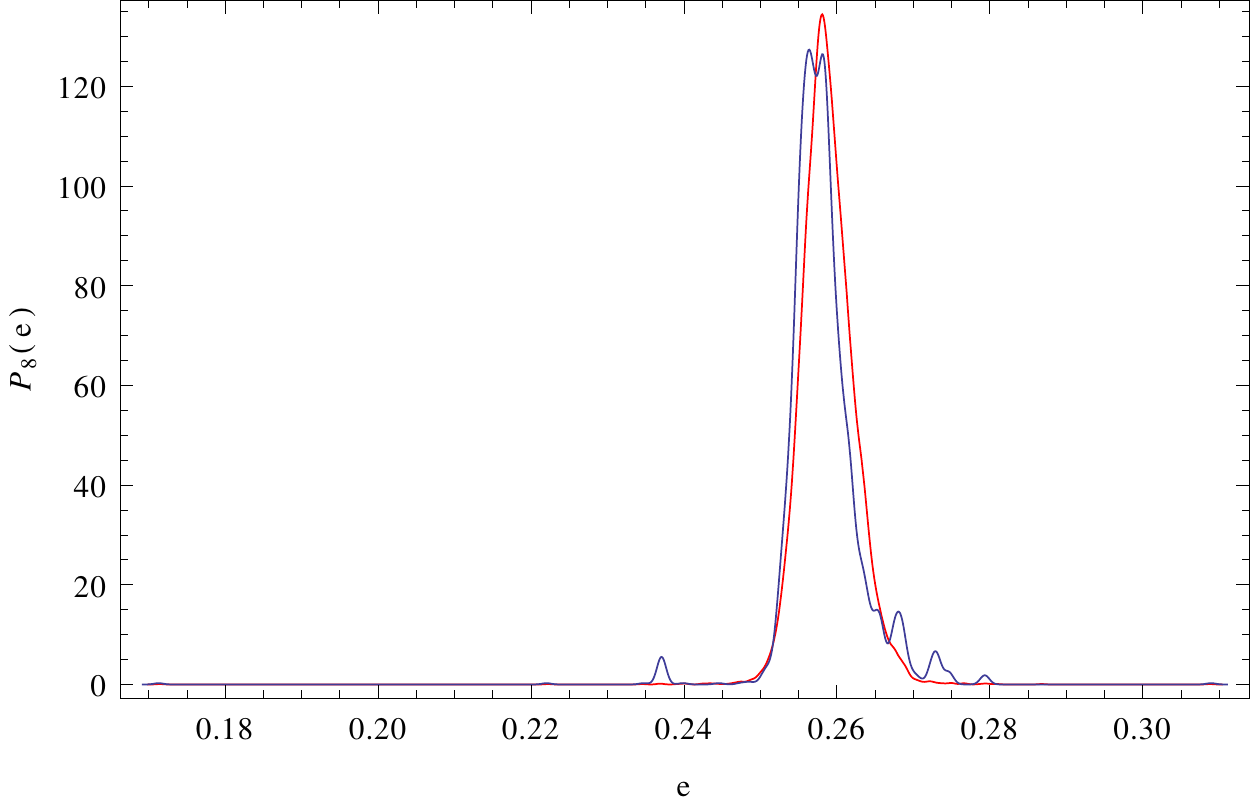}
\includegraphics[width=0.9\columnwidth]{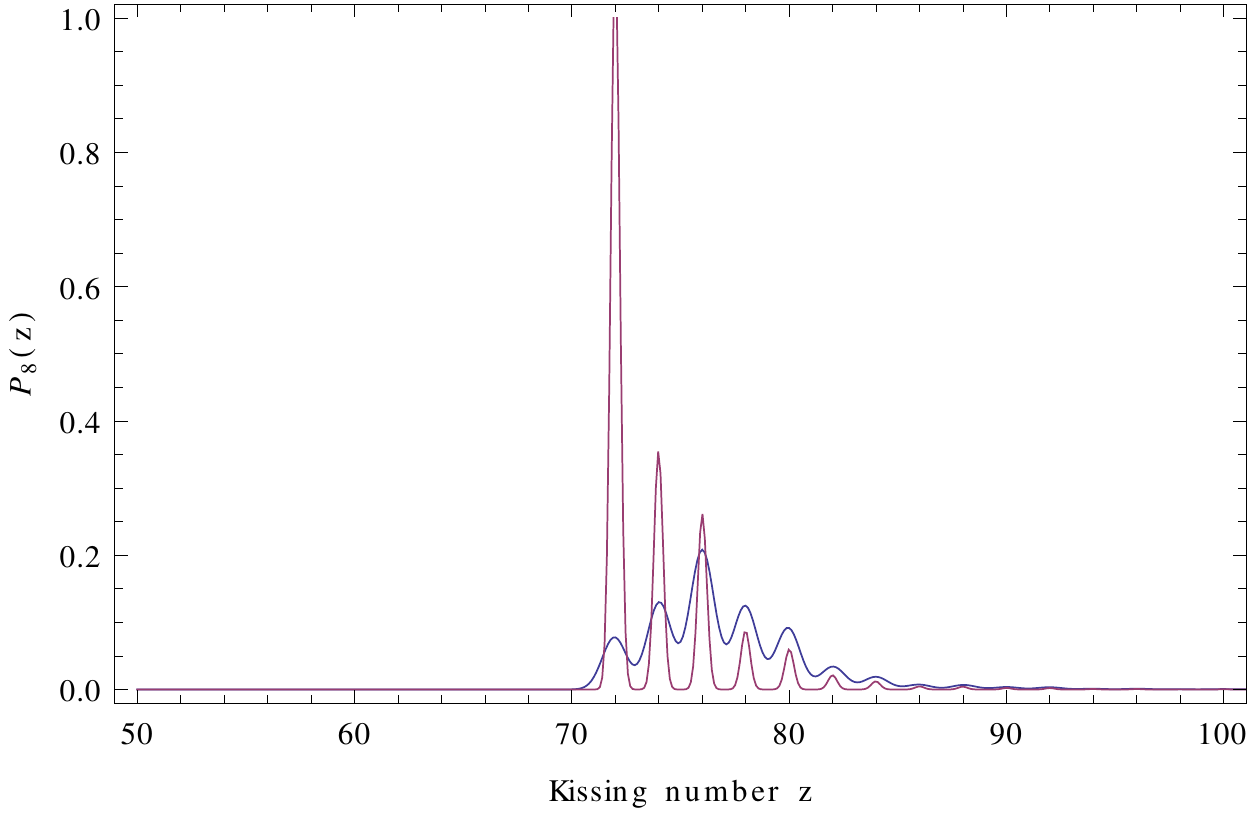}
\caption{Comparison of exact and empirical distributions generated by randomised Voronoi's algorithm. \emph{Top} Distribution of energies $e$ from randomised Voronoi's algorithm with isometry testing (blue) and exact distribution (red) for $d=8$. \emph{Bottom} Same comparison of distributions of kissing numbers from randomised Voronoi's algorithm with isometry testing (blue) and exact distribution (red) for $d=8$.}
\label{fig:exact-isom-d8}
\end{center}
\end{figure}
We see a reasonable agreement between the exact data and the ones generated by the randomised algorithm. This allows us to assume that data generated by randomised Voronoi's algorithm are representative and unbiased, and use data generated in higher dimensions where no exact data are available. The discrepancies present can be attributed to fluctuations associated to stochastic nature of our algorithm. This is especially clear for the kissing number which is integer by definition.

A rough measure of representativity of a sample generated by a random walk is whether it visits "dense" lattices with high kissing numbers, or even better - the densest (known) lattice in that dimension. For low dimensions, $d<13$, just $N_d=10^4$ runs were enough to satisfy this requirement. Starting with $d=13$ one has to make more runs (Although in $d=13$ a random walk of $10^4$ steps comes quite close to the best packing: $e=0.28$ and $e_\text{best}=0.27$). The required number of steps $N_d$ is growing fast: $N_{13}\sim10^5$, while $N_{14}>10^5$. The situation quickly deteriorates in higher dimensions: while in $d=8$ random walk is hitting $E_8$ around $80\%$ of the time, the number drops down to $<1\%$ of hits for $\Lambda_{10}$ - the best known lattice pakcing in $d=10$ - and goes further down for higher $d$. Table~\ref{fig:freq8-12} gives frequencies for a random walk to visit the best packers in $d=8-12$. The data seem to suggest a faster than exponential decay, a simple fit giving $\sim e^{-7.0\ x^{1.92}}$.
\begin{table}[htbp]
\begin{tabular}{ | c | c | c | c | c | c | }
\hline
Dimension & 8 & 9 & 10 & 11 & 12 \\
\hline
Frequency & 0.835 & 0.341 & 0.096 & 0.0156 & 0.00191\\
\hline
\end{tabular}
\caption{Frequencies with which a best known packing is visited by a random walk as a function of dimension.}
\label{fig:freq8-12}
\end{table}

\begin{figure}[htbp]
\begin{center}
\includegraphics[width=0.9\columnwidth]{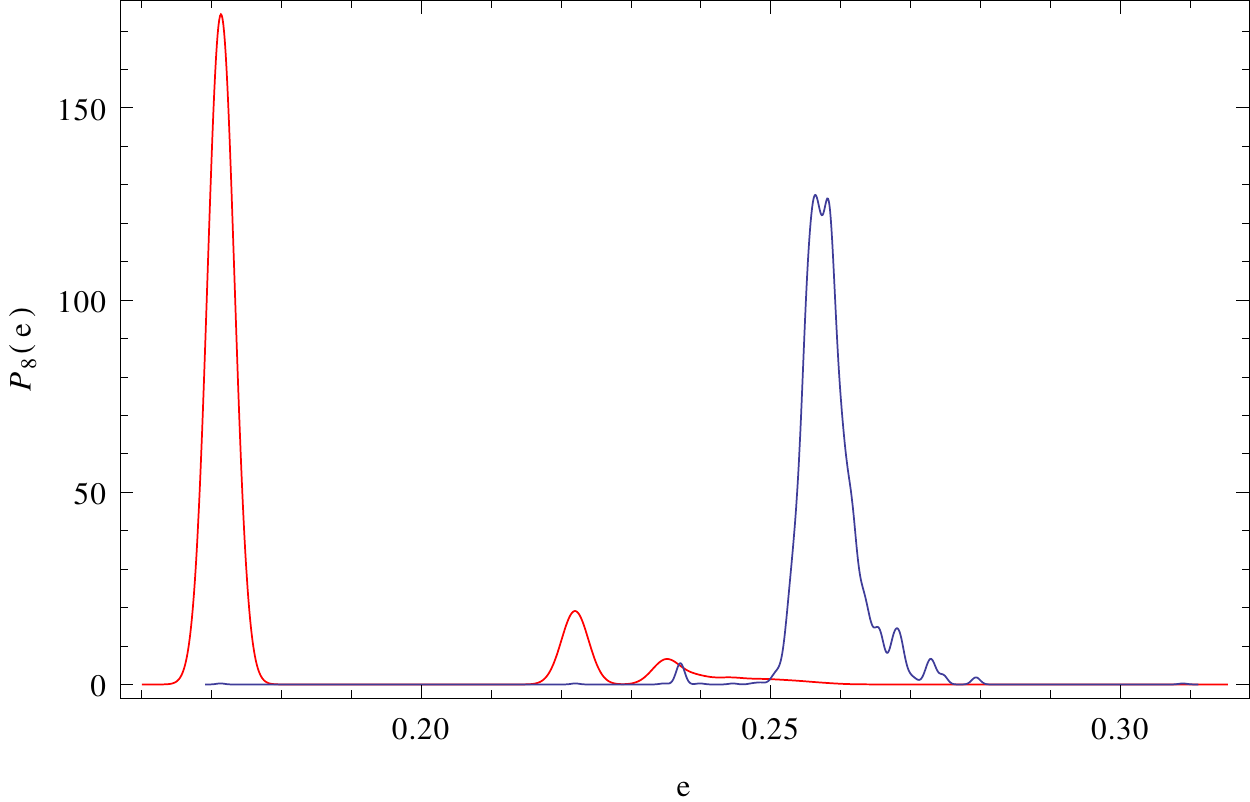}
\includegraphics[width=0.9\columnwidth]{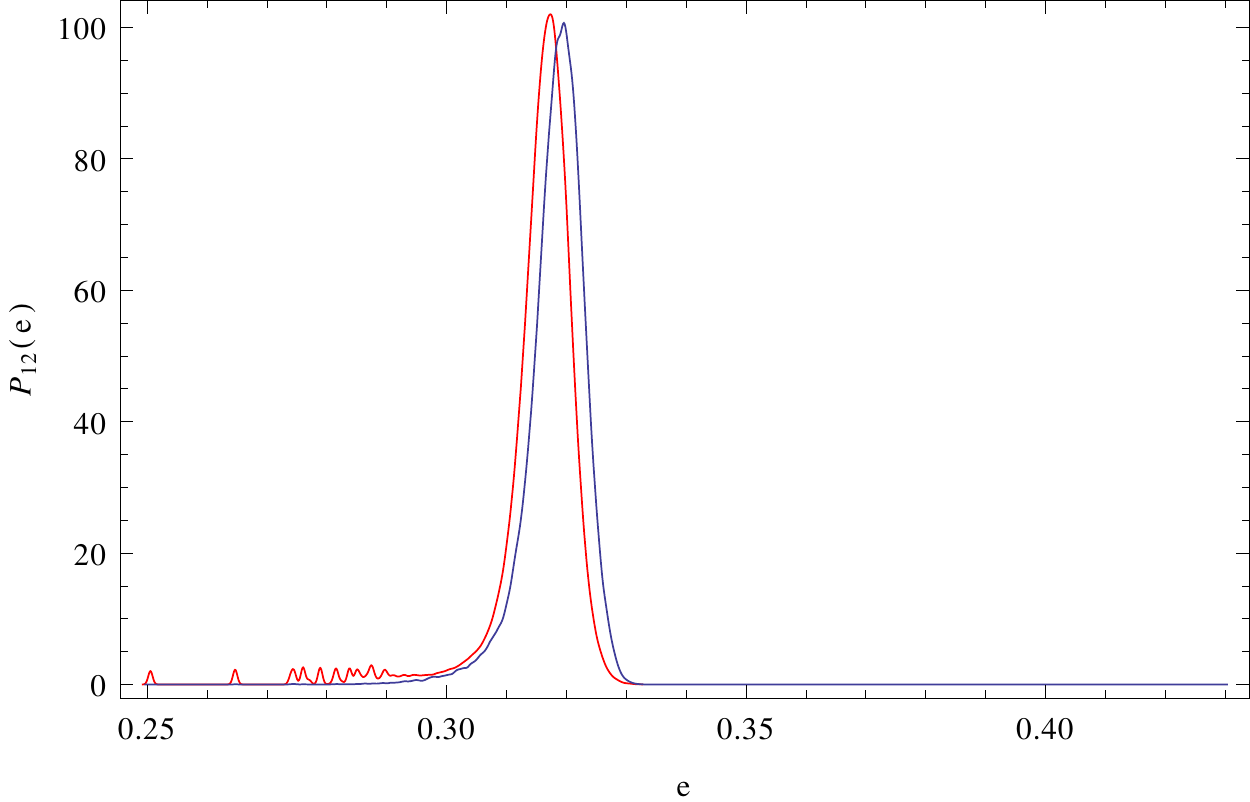}
\caption{\emph{Top. } Distribution of energies $e$ with (blue) and without (red) isometry test in $d=8$. Isometry is very important and the two distributions are completely different: without isometry check distribution concentrates around the energy of $E_8$). \emph{Bottom.} Distribution of energies $e$ with (blue) and without (red) isometry test in $d=12$. Isometry is no longer important and the distribtuions are almost the same, except low energy tail, where one still sees small spikes.}
\label{fig:isom-noisom}
\end{center}
\end{figure}

\begin{figure}[htbp]
\begin{center}
\includegraphics[width=0.9\columnwidth]{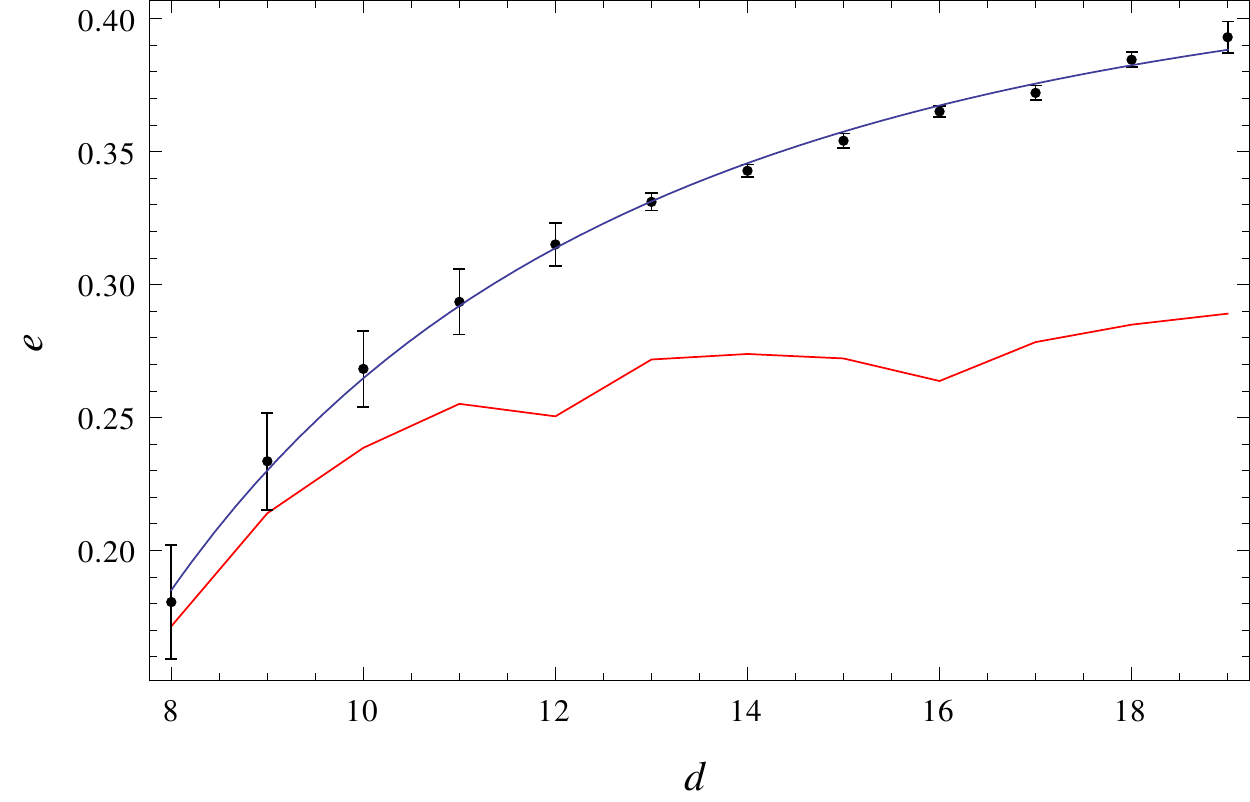}
\caption{Average energy $\langle e\rangle=-\frac{1}{d}\langle\log\varphi\rangle$ of a random walk as a function of dimension ($d=8-19$). Errorbars correspond to standard deviation of energy. The smooth curve is a guide to the eye.}
\label{fig:rwemm8to19}
\end{center}
\end{figure}

The table~\ref{tab:rwestat8to19} gives a summary on average energies, their standard deviations $\sigma_e$, best found, worst found and best known lattice for $d=8-19$ ($N$ is number of steps in random walk):
\begin{table}
\begin{tabular}{ | c | c | c | c | c | c | c | }
\hline
Dim. & $\mean{e}$ & $\sigma_e$ & Best found & Worst found & Best known \\
\hline
8 & 0.180572 & 0.021502 & 0.171465 & 0.308792 & 0.171465 \\
\hline
9 & 0.23352 & 0.01823 & 0.21396 & 0.34188 & 0.21396 \\
\hline
10 & 0.26828 & 0.01422 & 0.23857 & 0.37285 & 0.23857 \\
\hline
11 & 0.29347 & 0.01228 & 0.25511 & 0.40193 & 0.25511 \\
\hline
12 & 0.31505 & 0.00796 & 0.25055 & 0.38024 & 0.25041 \\
\hline
13 & 0.33106 & 0.00328 & 0.27179 & 0.40709 & 0.27178 \\
\hline
14 & 0.34277 & 0.00236 & 0.31862 & 0.43265 & 0.27386 \\
\hline
15 & 0.35405 & 0.00273 & 0.33522 & 0.45703 & 0.27218 \\
\hline
16 & 0.36507 & 0.00197 & 0.34235 & 0.48031 & 0.26370 \\
\hline
17 & 0.37205 & 0.00280 & 0.33949 & 0.50258 & 0.27833 \\
\hline
18 & 0.38322 & 0.00235 & 0.37805 & 0.39238 & 0.28489 \\
\hline
19 & 0.39000 & 0.00391 & 0.37909 & 0.40146 & 0.28903 \\
\hline
\end{tabular}
\caption{Average energies of perfect lattices for $d=8\cdots19$. Sample sizes $N$ are $10^6$ for $d=8-12$, $10^5$ for $d=13-16$, $2\cdot10^5$ for $d=17,18$ and $1.5\cdot10^5$ for $d=19$. The observed increase of standard deviation $\sigma_e$ for $d>17$ indicates that sample size was not big enough. Increasing the sample size decreases the deviation.}
\label{tab:rwestat8to19}
\end{table}
Standard deviation clearly decreases with dimensions; the increase for $d=17-19$ indicates that more runs are required to get a representative set of lattices. Indeed comparing behaviour of the deviation with number of runs for $d=17$ one sees the decrease as number of runs increases (the same behaviour is present in $d=18,19$):
\begin{table}
\begin{tabular}{| c | c | c | c |}
\hline
$N$ & $10^4$ & $10^5$ & $2\cdot10^5$ \\
\hline
Deviation & 0.0064997 & 0.0030565 & 0.0028058 \\
\hline
\end{tabular}
\caption{Standard deviation of energy in $d=17$ as a function of number of runs $N$.}
\end{table}
The decrease of standard deviation suggests that distribution of energies $\mathcal{P}_d(e)$ is concentrating around mean value and becomes peaked around its mean value for large $d$ and for $d=\infty$:
\begin{equation}
\mathcal{P}_{d\to\infty}(e)\sim\delta(e - \langle\,e\rangle_{d\to\infty}).
\end{equation}
Fig.~\ref{fig:rwemm8to19} shows behaviour of average energy (no checks for isometry) with dimension. Large deviations in low dimensions up to $d<12$, represented by errorbars on the figure, are related to the fact that the distribution of energies in these dimensions is highly irregular if no check for isometry is performed during the random walk (see Fig.~\ref{fig:isom-noisom}, case of $d=8$ for an illustration).

An important issue is equivalence/isometry of generated lattices. As we have discussed above a single lattice admits many equivalent representations in terms of quadratic forms. One might worry if random walk is generating many/few equivalent lattices. The above results were generated neglecting isometry partially: ony LLL-reduction was performed on newly generated forms. Based on $d=7,8$ results we know that isometry is definitely important in low dimensions. However it is only relevant for low dimensions, our data suggest $d<13$, where the number of perfect lattices is relatively small and random walks of moderate size contain many isometric copies of the same lattice. For higher dimensions, $d\geq13$, where the number of perfect lattices is huge the chance of hitting an isometric lattice is vanishingly small except for the densest lattices which have a larger isometry family. This is illustrated by Fig~\ref{fig:isom-noisom} which compares probability distributions of energies for $d=8$ and $d=12$.
It is worth stressing that this statement holds true only if one samples a relatively small subset of all perfect lattices. Once sample size is comparable to the size of the full set of perfect forms, isometry becomes important in any dimension. This fact can in principle be used to define a formal criterium whether one has generated a respresentative sample. Including isometry test in generation procedure is easy: every newly generated form  is checked for isometry against all previously genrated forms~\footnote{See Appendix B for more details}. 



The effect of isometry on energy average $\langle\,e\rangle$ is to increase values for low dimensions, which are dominated by dense lattices if no isometry cheks are performed. The higher-dimensional data, $d>12$ are left intact since isometry becomes completely irrelevant. We reproduce the table~\ref{tab:rwestats8to12} 
and the $\langle\,e\rangle$ curves for data with isometry checks. We see the same trend of decreasing standard deviation with increase of dimension as in the case of no isometry testing. In what follows we are using samples with checks for isometry for $d<12$ and with no isometry checks for $d>11$.
\begin{table}
\begin{tabular}{ | c | c | c | c | c | c | c | }
\hline
$d$ & $\langle e\rangle$ & $\text{Std}(e)$ & $\langle e\rangle_i$ & $\text{Std}_i(e)$ & $\langle e\rangle_\text{ex}$ & $\text{Std}_\text{ex}(e)$\\
\hline
8 & 0.180571 & 0.021502 & 0.258296 & 0.0050364 & 0.258845 & 0.003593 \\
\hline
9 & 0.233521 & 0.018231 & 0.266341 & 0.005073 & $0.259662^*$ & $0.006006^*$ \\
\hline
10 & 0.268281 & 0.014227 & 0.281615 & 0.005484 & - & - \\
\hline
11 & 0.293471 & 0.012288 & 0.299142 & 0.005262 & - & - \\
\hline
12 & 0.31506 & 0.007967 & - & - & - & - \\
\hline
\end{tabular}
\caption{Comparison of energy averages $\langle\,e\rangle$ without and with isometry test. Additionaly exact values of average and standard deviation are given for $d=8$. $^*\text{V}$alues for $d=9$ are extracted from partial enumeration~\cite{schurmann2012perf9}.}
\label{tab:rwestats8to12}
\end{table}

In what follows we use mixed set of data: samples with isometry checks for $d<13$ and samples with no isometry testing applied for $d>12$. We do so to remove features specific to low dimensions $d<13$ and reveal the generic features common with dimensions $d>12$.

\begin{figure}[htbp]
\begin{center}
\includegraphics[width=0.9\columnwidth]{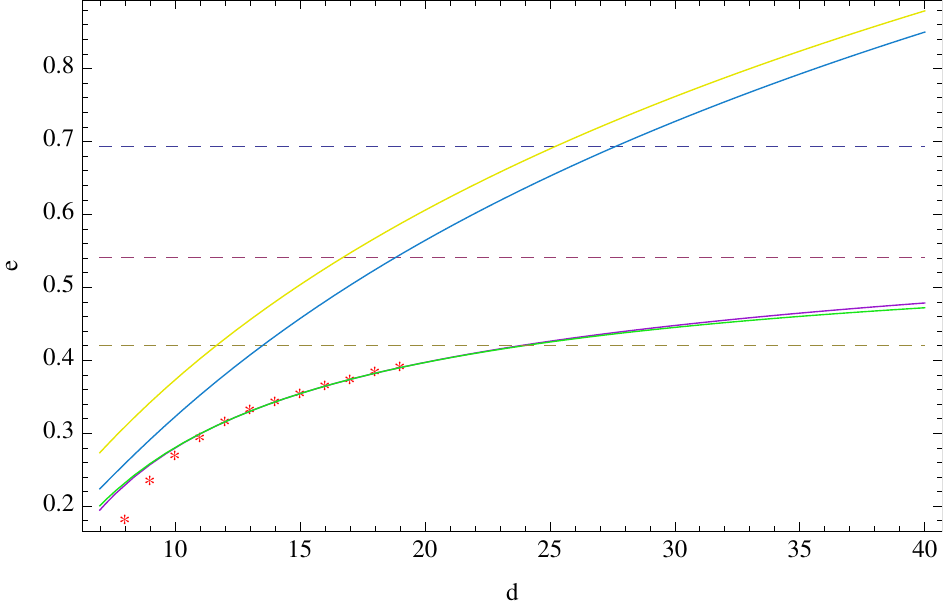}
\label{fig:rwemmcmp8to19}
\caption{Average energy $\langle e\rangle=-\frac{1}{d}\langle\log\varphi\rangle$ of a random walk as a function of dimension ($d=8-19$) (red crosses) compared to energy of $A_d$ and $D_d$ lattices (see~Eq.\eqref{eq:eaddd}). The dashed curves are the leasing asymptotics of Minkowksy (top), Torquato-Stillinger (middle) and Kabatiansky-Levenstein (bottom) bounds. The Minkowksy and  Torquato-Stillinger are upper bounds while the Kabatiansky-Levenstein bound is a lower bound on the energy of the best packing. The yellow and blue continuous lines are $A_d$ and $D_d$, the green and violet lines are the two fits (\ref{eq:fitcons}), (\ref{eq:fitlog}).}
\end{center}
\end{figure}

Let us concentrate on two possible scenarios, the simplest cases where to locate our typical lattices. On one hand we can for example look at the energies of $A_d$, $D_d$ families of lattices\cite{zachary2011high}:
\begin{eqnarray}
\label{eq:eaddd}
e(A_d) & = & \frac{1}{2}\log\frac{2}{\pi} + \dfrac{\log(1+d)}{2d} + \frac{1}{d}\log\Gamma(1+\frac{d}{2})\\
&\simeq& \log(d)/2+\Ord{1}\\
e(D_d) & = & -\frac{1}{2}\log\,\pi + (\frac{1}{2}+\frac{1}{d})\log\,2 + \frac{1}{d}\log\Gamma(1+\frac{d}{2})\\
&\simeq& \log(d)/2+\Ord{1}\notag
\end{eqnarray}
Both $A_d$ and $D_d$ have asymptotically equal energies for $d\to\infty$: $\sim\log d/2$ which means sub-exponential packing fraction. 

Minkosky's and Kabatiansky-Levenstein bounds tell us that there are lattices with only exponentially small packing fraction. Asymptotically in large dimensions, upper and lower bounds give:
\begin{eqnarray}
e_{M} &=& \log(2)+\Ord{\log(d)/d}\\
e_{KL} &=&0.413...
\end{eqnarray}
and it is worth remembering the Torquato-Stillinger conjectured bound which should replace Minkoswky's under appropriate hypothesis on high-dimensional lattices \cite{torquato2006new, scardicchio2008estimates}:
\begin{equation}
e_{TS}=0.539+\Ord{\log(d)/d}.
\end{equation}

Random walks in high dimensions are sampling lattices with energy close to its mean value $\mean{e}$. We try two fits for this function of $d$, one with the leading order term constant, hypothesizing a ``best packer" behavior for typical lattices in high dimensions and the other with leading $\log(d)$\footnote{We use 8 points between $d=12$ and $d=19$, no sensible differences are obtained including less points in this range.}. For the first we obtain
\begin{equation}
\label{eq:fitcons}
\mean{e}=(0.58\pm 0.04)-\frac{\log(d)}{d}(0.9\pm1.0)-(0.8\pm0.6)d^{-1}.
\end{equation}
The constant term is suggestively close to the Torquato-Stillinger bound and, within the associated error, it is below the Minkowsky bound $\log(2)=0.69$. However, an equally good fit can be obtained by assuming that the leading term is growing logarithmically
\begin{equation}
\label{eq:fitlog}
\mean{e}=(0.066\pm 0.04)\log(d)+(0.27\pm 0.04)-(1.4\pm 0.2)d^{-1}
\end{equation}
although the coefficient of the logarithm is well below the value $0.5$ of the $A_d$ and $D_d$ families (typical lattices are much denser than these examples). Both fits are equally good, as can be seen from Fig.~\ref{fig:rwemm8to19}, the resolution of the two can only occur for $d\gg 40$.



\begin{figure}[htbp]
\begin{center}
\includegraphics[width=0.45\columnwidth]{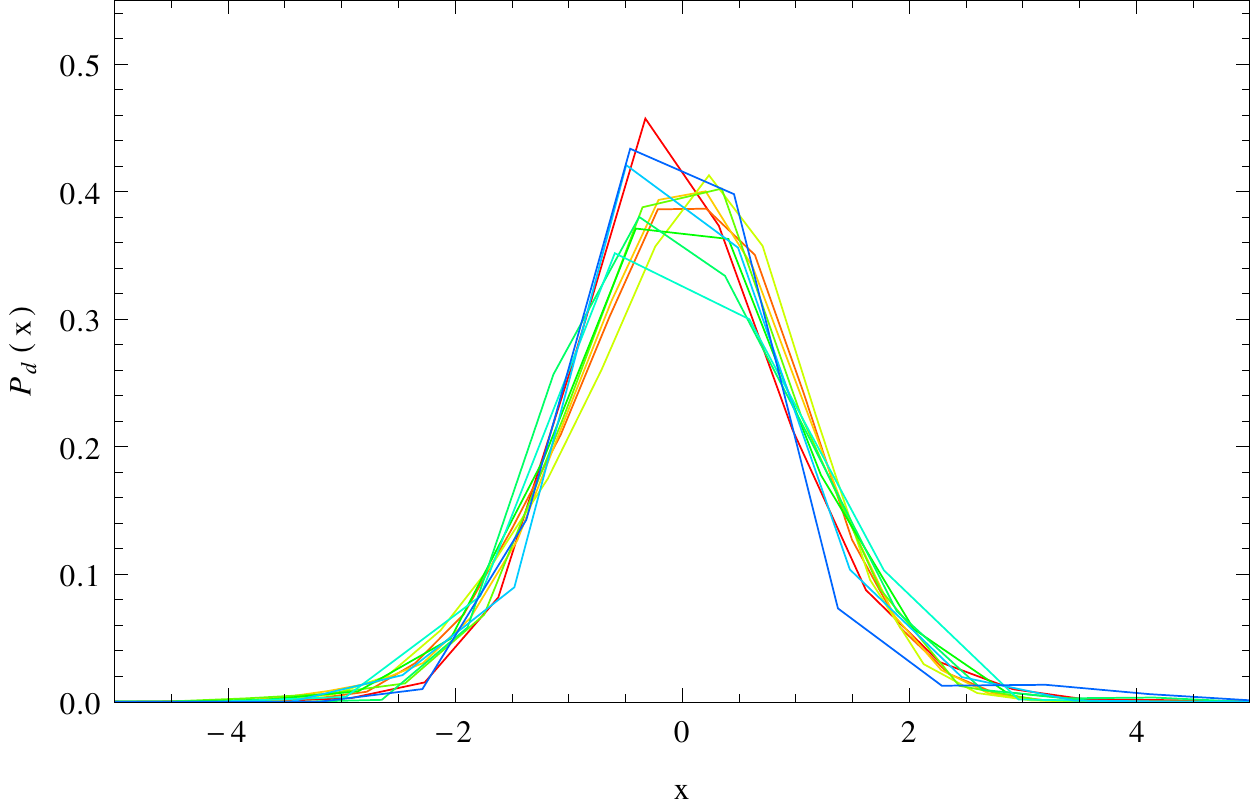}
\includegraphics[width=0.45\columnwidth]{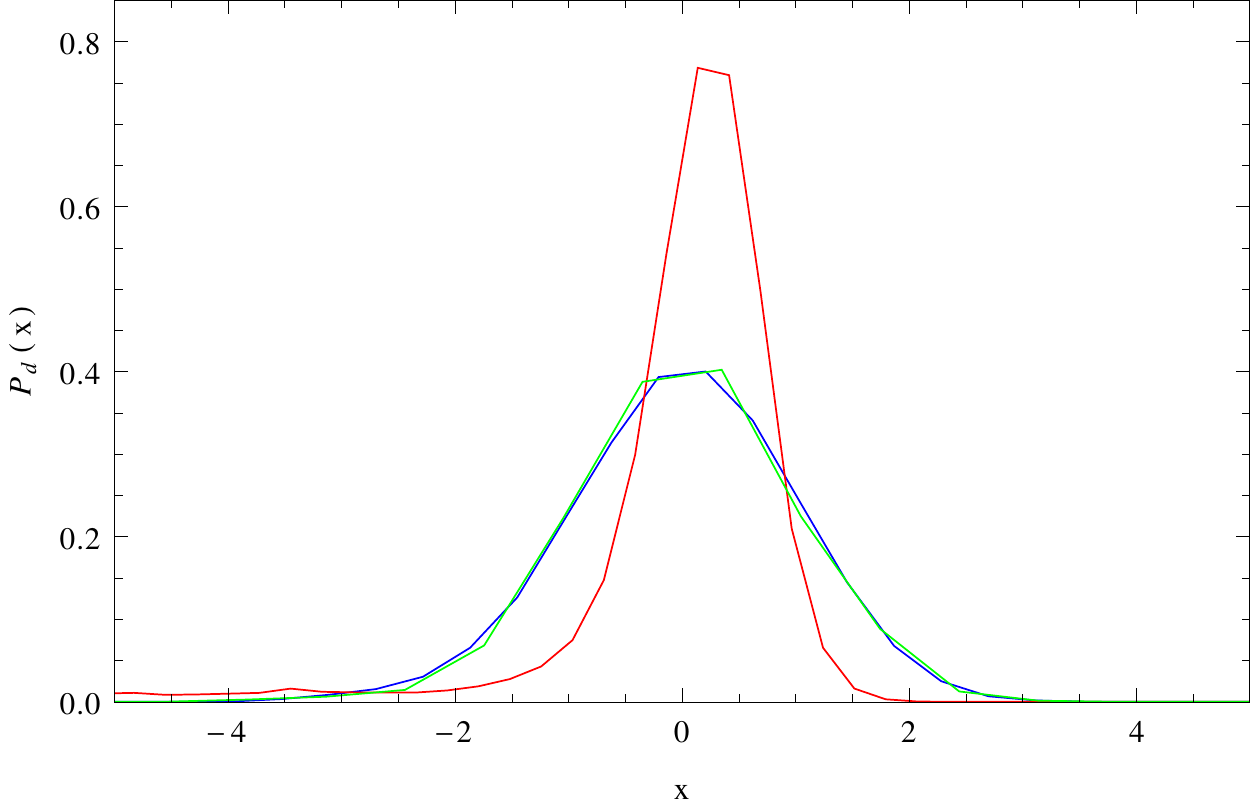}
\caption{\emph{Left} Probability distributions $\mathcal{P}(x)$ for $d=8-10,13-17$ - color goes from red for $d=8$ to magenta for $d=19$. We have used exact distribution for $d=8$ for convenience and skipped $d=12$. \emph{Right} Comparison of distributions $\mathcal{P}_d(x)$ for $d=10,12,13$. The case $d=12$ is very disctinct from neighbouring  dimensions.}
\label{fig:rwpdfx8to19}
\end{center}
\end{figure}

\begin{figure}[htbp]
\begin{center}
\includegraphics[width=0.9\columnwidth]{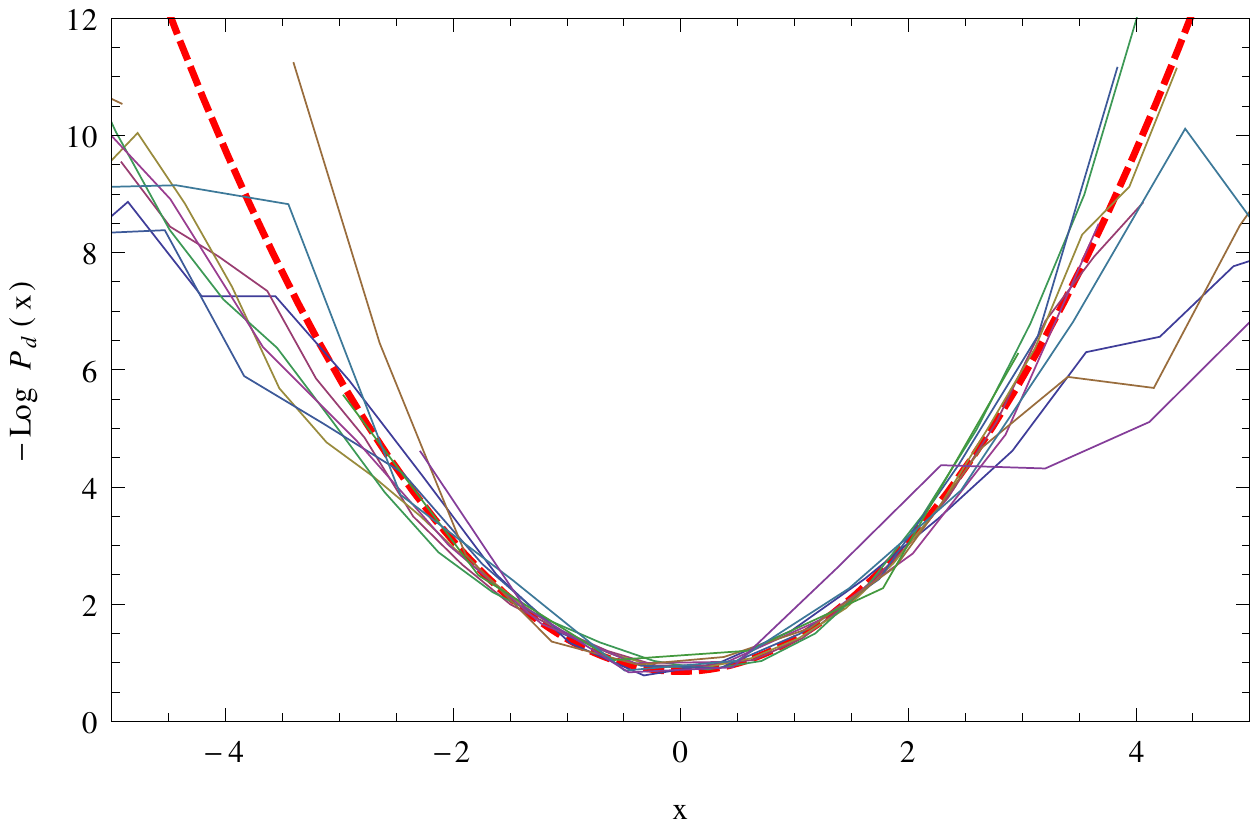}
\caption{Gaussian fit to the central part of the probability ditribution $\mathcal{P}_d(x)$ for $d=8-11,13-19$.}
\label{fig:rwpdflx8to19}
\end{center}
\end{figure}

The main effect of isometry on distribution of energies $\mathcal{P}(e)$ is to supress low energy spikes (see Fig.~\ref{fig:isom-noisom}) associated with dense lattices which are relatively often visited in these dimensions by a random walk, and shift the weight to the universal bell-like feature which dominates the distribution $\mathcal{P}_d(e)$ in high dimensions. As of the distribution of kissing numbers $Z$ switching on the isometry testing kills the large-$Z$ tail of the distribution and concentrates the weight around small values of $Z$ of order $d(d+1)$ (recall that this is the lower bound on kissing number for perfect lattices). These facts indicate that in high dimensions typical perfect lattices have relatively high energy (but still lower than $A_d$ and $D_d$) and small kissing numbers, of order $d(d+1)$.
\begin{figure}[htbp]
\begin{center}
\includegraphics[width=0.9\columnwidth]{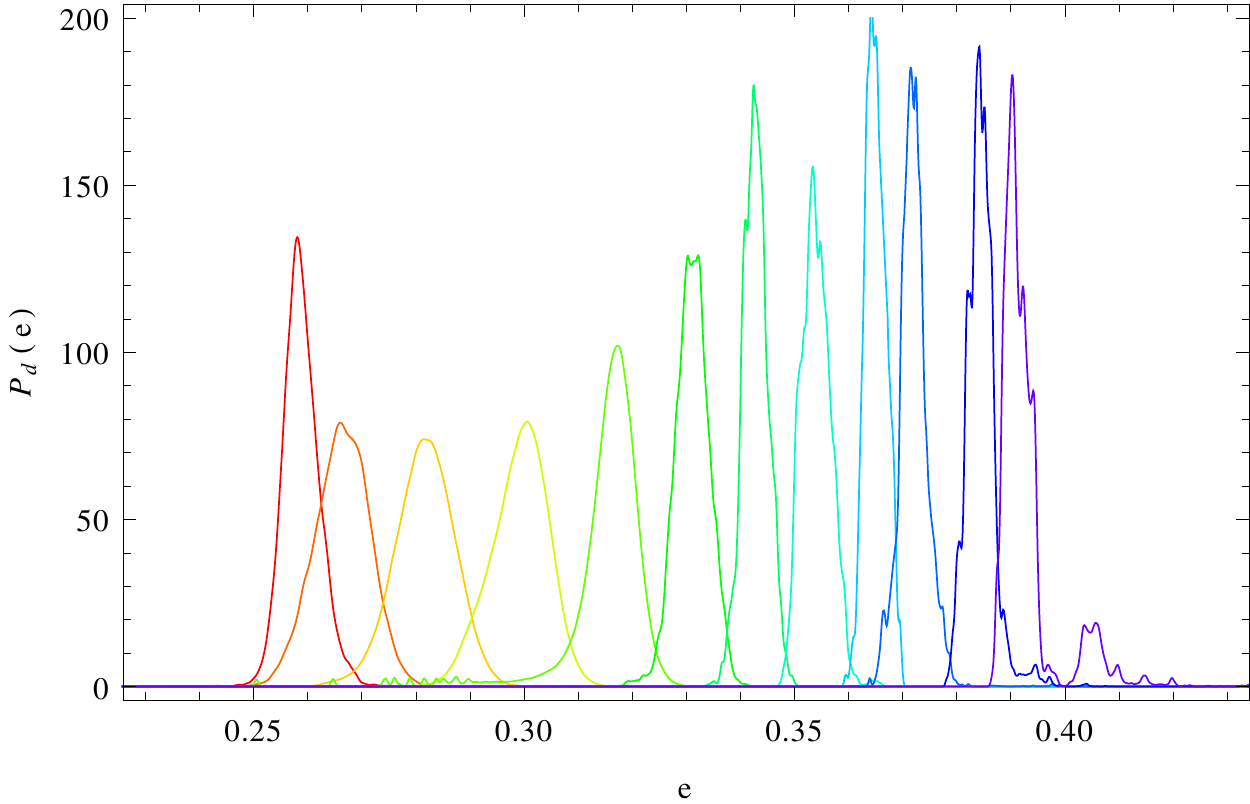}
\caption{Probability distributions $\mathcal{P}_d$ of energy $e$ for $d=8-10,12-19$ - color goes from red ($d=8$) to violet ($d=19$). As dimension increases averages increase and peaks shift to the right.}
\label{fig:rwpdfe8to19}
\end{center}
\end{figure}
If we define rescaled variable $x=(e-\langle\,e\rangle_d)/\sigma_e$ we expect the probability distribution functions of $x$  to collapse on some master curve with mild dependence on $d$: $$\mathcal{P}_d(x)\propto\mathcal{P}_d\left(\dfrac{e-\langle\,e\rangle}{\sigma_e}\right).$$ Indeed after rescaling a master curve is emerging as shown on Fig.~\ref{fig:rwpdfx8to19} though the collapse is not perfect: case $d=12$ is special with quite different shape as compared to other dimensions as highlighted on Fig.~\ref{fig:rwpdfx8to19}. All the distributions are skewed to the left, i.e. towards denser lattices, although this is hard to spot on Fig.~\ref{fig:rwpdfx8to19} while this is clearly so for $d=12$. These features become more pronounced if one studies $g_d(x)=-\log\,\mathcal{P}_d(x)$ showed on Fig.~\ref{fig:rwpdflx8to19}: the generic skeweness to the left (towards the denser lattices) becomes clear. For all dimensions studied except $d=12$ the central part of $g_d(x)$ can be well fited with a Gaussian $$-\log\mathcal{P}_d(x\sim0)\sim 0.85 + \dfrac{x^2}{1.8},$$ the value of the coefficient of $x^2$ being slightly larger than (but still consistent with) $1/2$ reflects the skewness of the distribution. The skewness only appears for larger values of $x$ which are noisy because we do not have enough statistics to probe them accurately.

\begin{figure}[htbp]
\begin{center}
\includegraphics[width=0.9\columnwidth]{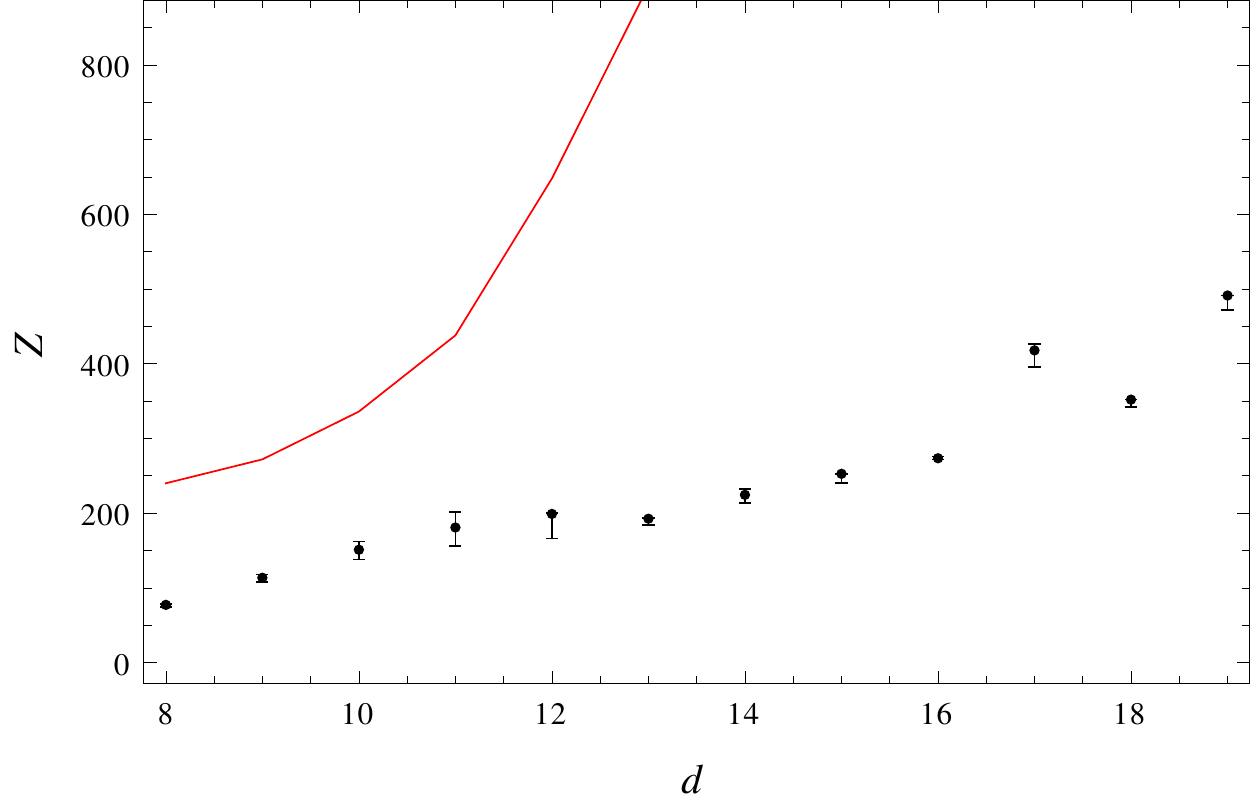}
\caption{Average kissing number for $d=8-10,12-19$. The red curve is the best known kissing numbers in corresponding dimensions.}
\label{fig:rwzm8to19}
\end{center}
\end{figure}

\begin{figure}[htbp]
\begin{center}
\includegraphics[width=0.9\columnwidth]{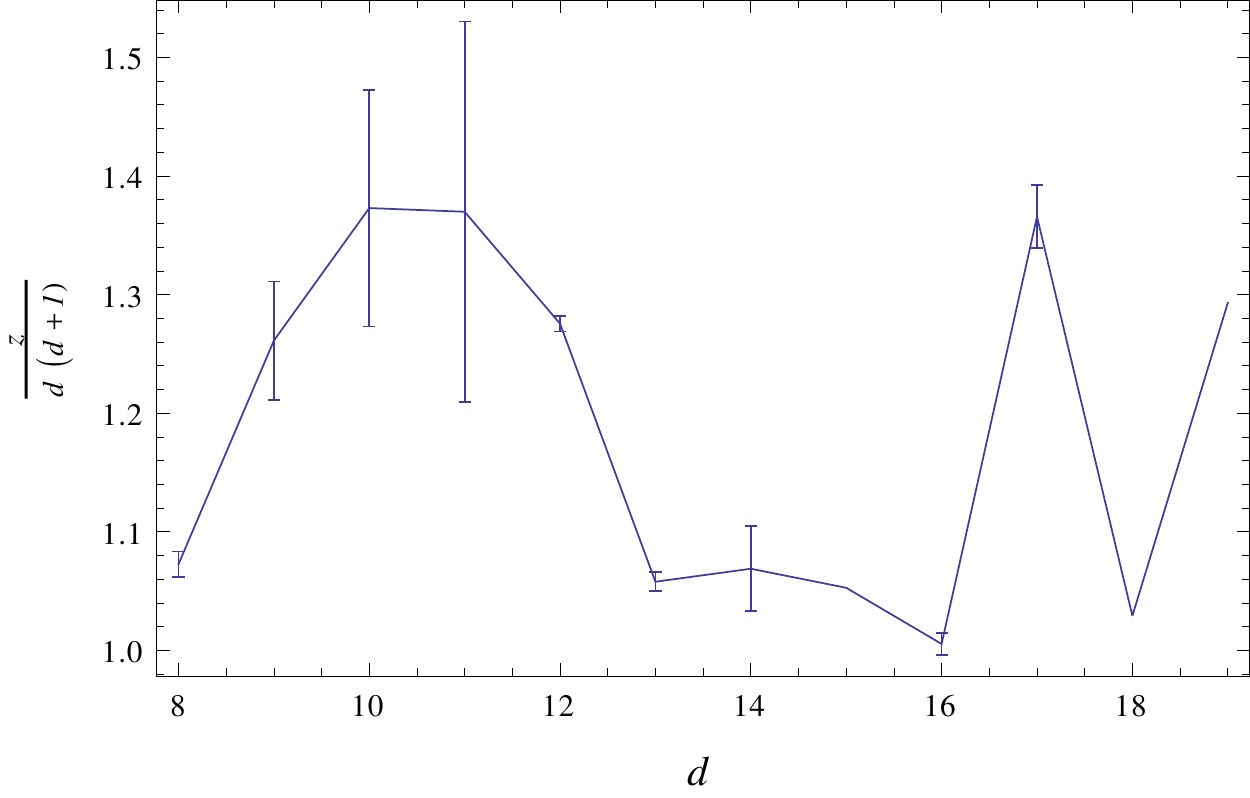}
\caption{Average kissing number normalised by $d(d+1)$ for $d=8-19$. Errorbars correspond to first and third quartiles (These are zero for $d=18,19$). Despite strong fluctuations the value of normalised kissing numbers is of order $1$.}
\label{fig:rwtpzm8to19}
\end{center}
\end{figure}

We now study the statistics of kissing number. For a typical perfect lattice the kissing number is of order $d^2$, i.e.\  like for $A_d$ or $D_d$, and of the same order of magnitude as the lower bound $d(d+1)$. To highlight this point we normalized $\mean{z}$ by $d(d+1)$, the minimal possible kissing number which gave a curve shown on Fig.~\ref{fig:rwtpzm8to19}. Thus a typical perfect lattice is similar to $A_d$ or $D_d$ in kissing numbers but has a lower energy/higher packing fraction. As we see from Figs.~\ref{fig:rwzm8to19} and \ref{fig:rwtpzm8to19} kissing number fluctuates much stronger than energy and the only conclusion we can make from the plots is that the distributions concentrate around their means just like it happens with energy. Combining this observation together with behavior of average energy we see that in high dimensions the Voronoi graph is dominated by lattices which have properties similar to $A_d$ and $D_d$.

\subsection{Random walk with $\beta>0$}

As dimension is increased beyond $d\sim 13$ we are no longer able to recover the densest known lattice packing with a plain random walk, at least for the number of steps we have tried (from a few hundred thousands to a few millions, depending on dimension). Given a fast growth of the number of perfect forms with dimension, one would likely have to sample random walks of size comparable to the number of perfect forms to see the densest lattices, something that is out of reach already for moderate dimensions $d\sim13-14$.

We therefore introduced a procedure which biases the walk towards denser lattices. We employed standard Metropolis-like rule with fictitious temperature $\beta$ described above in Sec.~\ref{sec:biasrwbeta} which favours denser lattices. Namely, we generate a neighbor $Q'$ of the lattice $Q$ and compute its packing fraction $\phi(Q')$ and from this its energy $e(Q')$. If $e(Q')\leq e(Q)$ we accept the move and if $e(Q')>e(Q)$ we accept the move only with probability $\exp(-\beta(e(Q')-e(Q)))$. 

This allowed us to recover consistently the densest (known) lattice packings up to $d=17$ and to get very close to the best known lattices in $d=18,19$, where we start seeing some complex landscape behavior. We managed to get the best known pakcing in these dimensions too but in a much less consistent fashion.

\begin{figure}[htbp]
\begin{center}
\includegraphics[width=0.9\columnwidth]{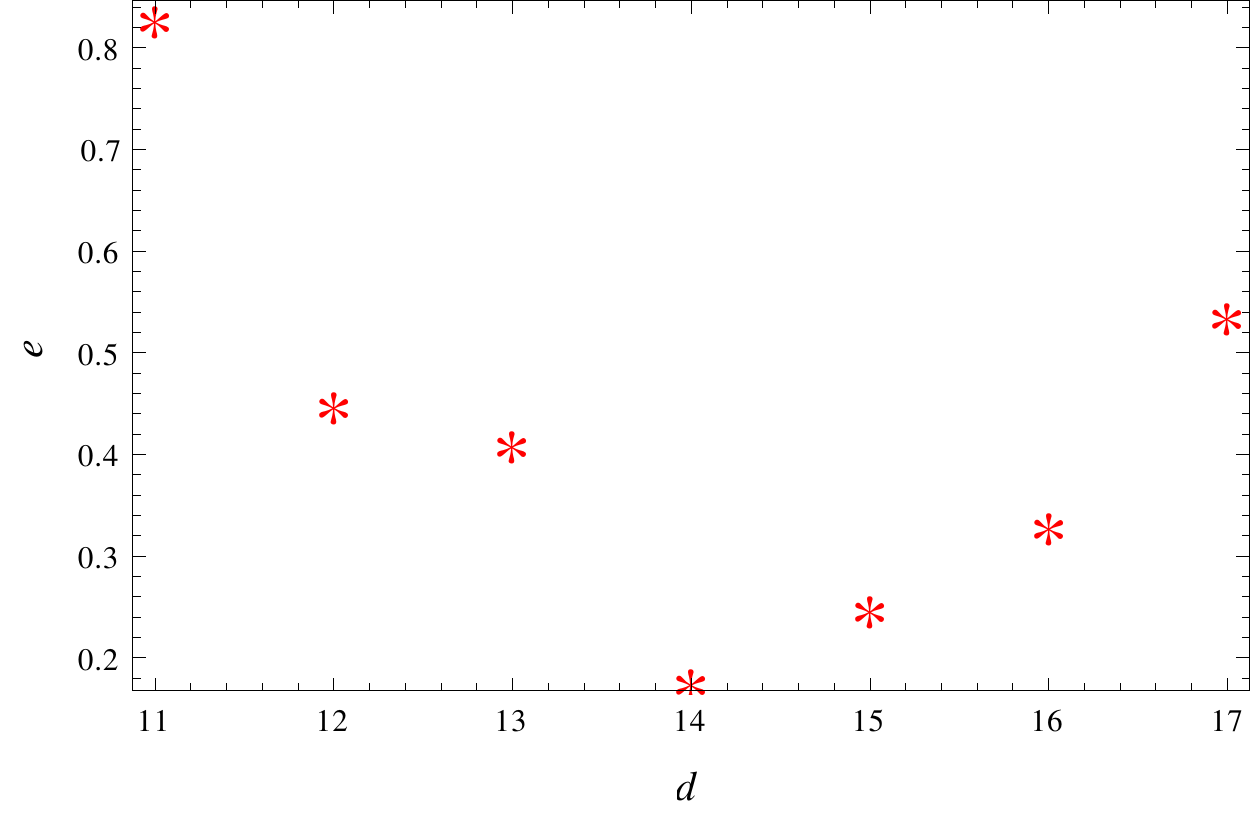}
\caption{$W=(E_>-E_<)/E_c$ as function of dimension $d=8-18$.}
\label{fig:mc8to18ebscales}
\end{center}
\end{figure}

Again we are looking at distributions and moments -- average and standard deviation -- of energy and kissing number. We saw for plain random walk which corresponds to $\beta=0$ that $E(d)=\langle\,e\rangle$ is a smooth curve as a function of dimension. As the temperature is lowered $E_\beta(d)$ curves become more singular reflecting the peculiarities of any given dimension: it is well known that the nature of dense sphere packings varies greatly as a function of dimension -- one of the factors that makes the problem of sphere packing so complicated.
\begin{figure}[htbp]
\begin{center}
\includegraphics[width=0.9\columnwidth]{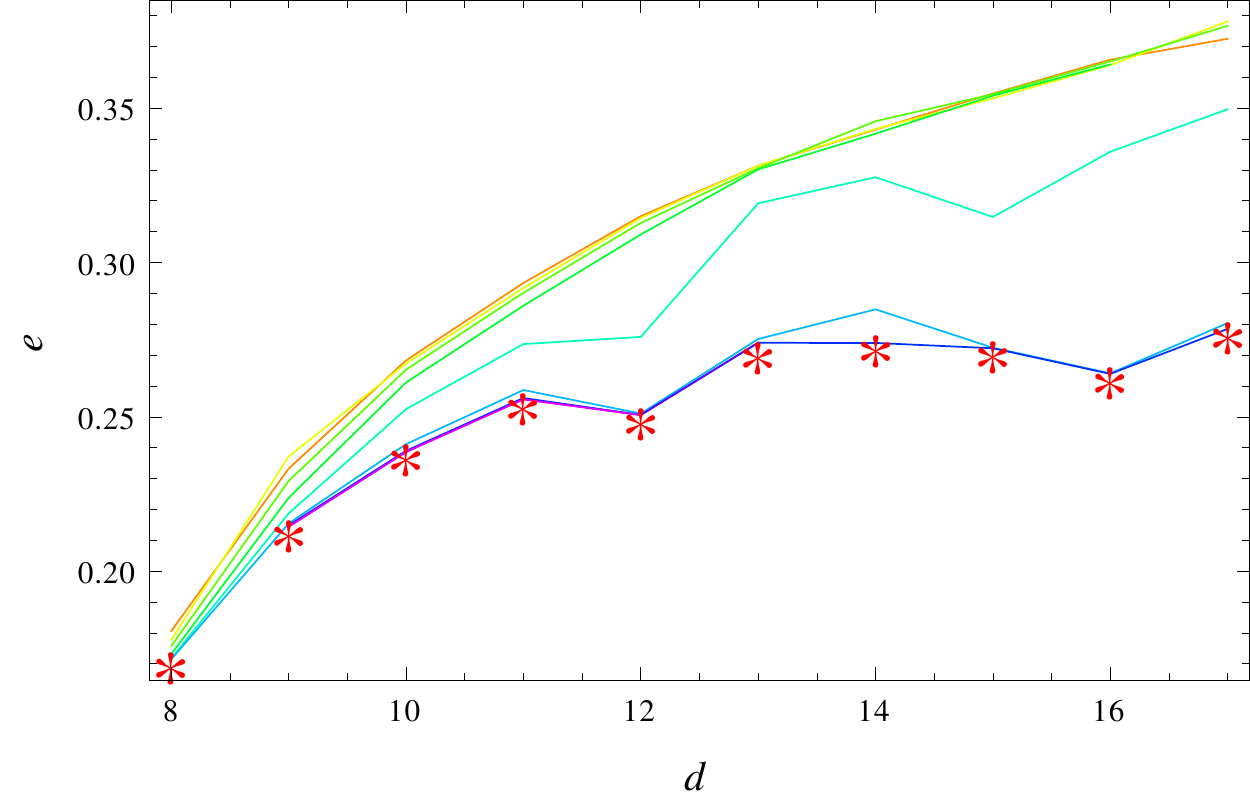}
\caption{\emph{Top} Average energy $\langle e\rangle=-\frac{1}{d}\langle\log\phi\rangle$ of a biased random walk as a function of dimension $d=8-17$: inverse temperature $\beta$ goes from $0$ (red) to $5$ (violet); red crosses are the best known lattice packings. As the temperature is decreased, details of the scenarios in finite dimensions become relevant.}
\label{fig:mc8to19ebbeta2}
\end{center}
\end{figure}
Up to $d=11$ changing the temperature immediately affects the range of energies probed by the random walk: the lower the temperature the lower the energy and $E(\beta)=\langle\,e\rangle_\beta$ is essentially an exponentially decaying function of $\beta$.
\begin{figure}[htbp]
\begin{center}
\includegraphics[width=0.9\columnwidth]{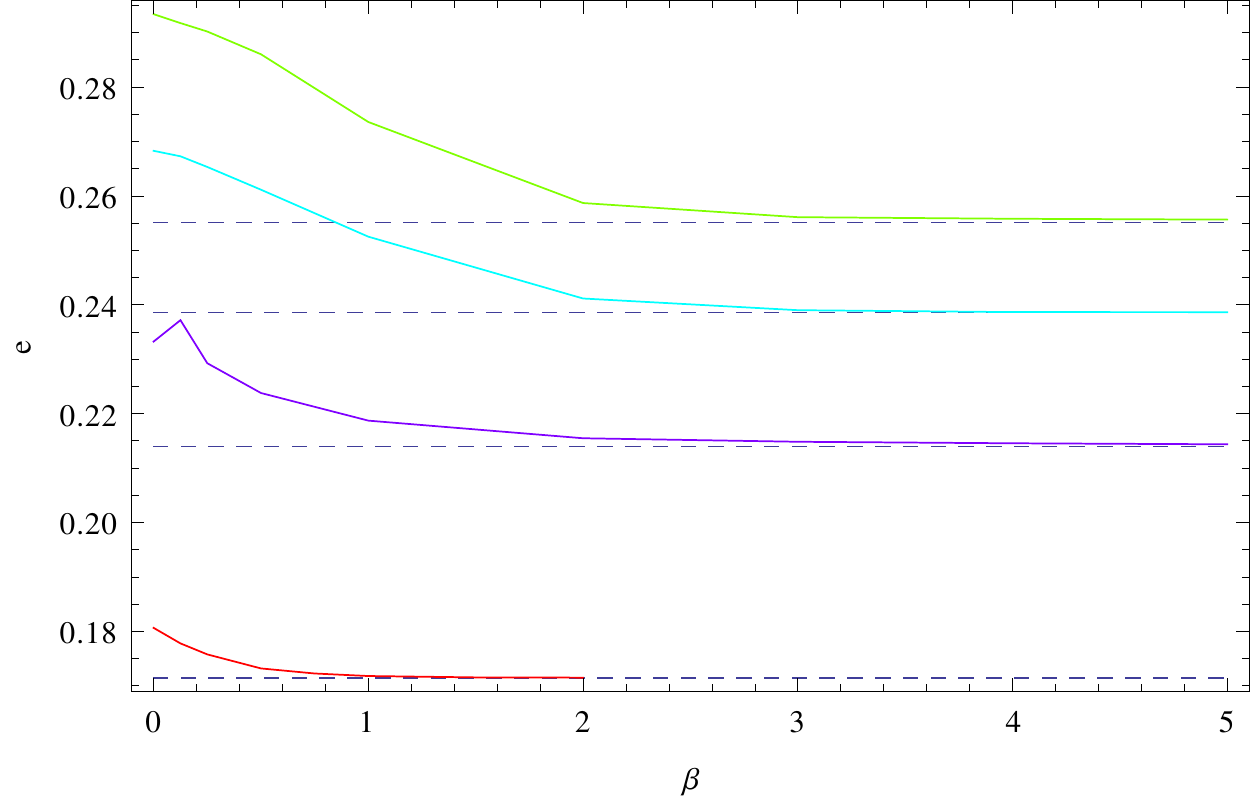}
\caption{Average energy $\langle e\rangle=-\frac{1}{d}\langle\log\phi\rangle$ of a biased random walk as a function of temperature for dimensions $d=8-11$ (color goes from red to green); dashed lines are the best known energies in corresponding dimensions.}
\label{fig:mc8to11ebbeta}
\end{center}
\end{figure}
Starting from $d=12$  and up the pattern of $E(\beta)$ changes qualitatively: a plateau emerges at small $\beta$ where the probed energy is almost insensitive to variations of temperature and is roughly equal to energy of $\beta=0$ random walk. As inverse temperature $\beta$ is increased there is a crossover to lower value of energy. The value $E(\beta)$ for large $\beta$ is approximately equal to the ground state energy, again almost insensitive to variation of $\beta$. Furthermore, sufficiently close to the crossover we observe strong run to run fluctuations of values of $\langle\,e\rangle_\beta$, a phenomenon which is reminiscent of a glassy free energy landscape \cite{mezard1987spin}. 
\begin{figure}[htbp]
\begin{center}
\includegraphics[width=0.9\columnwidth]{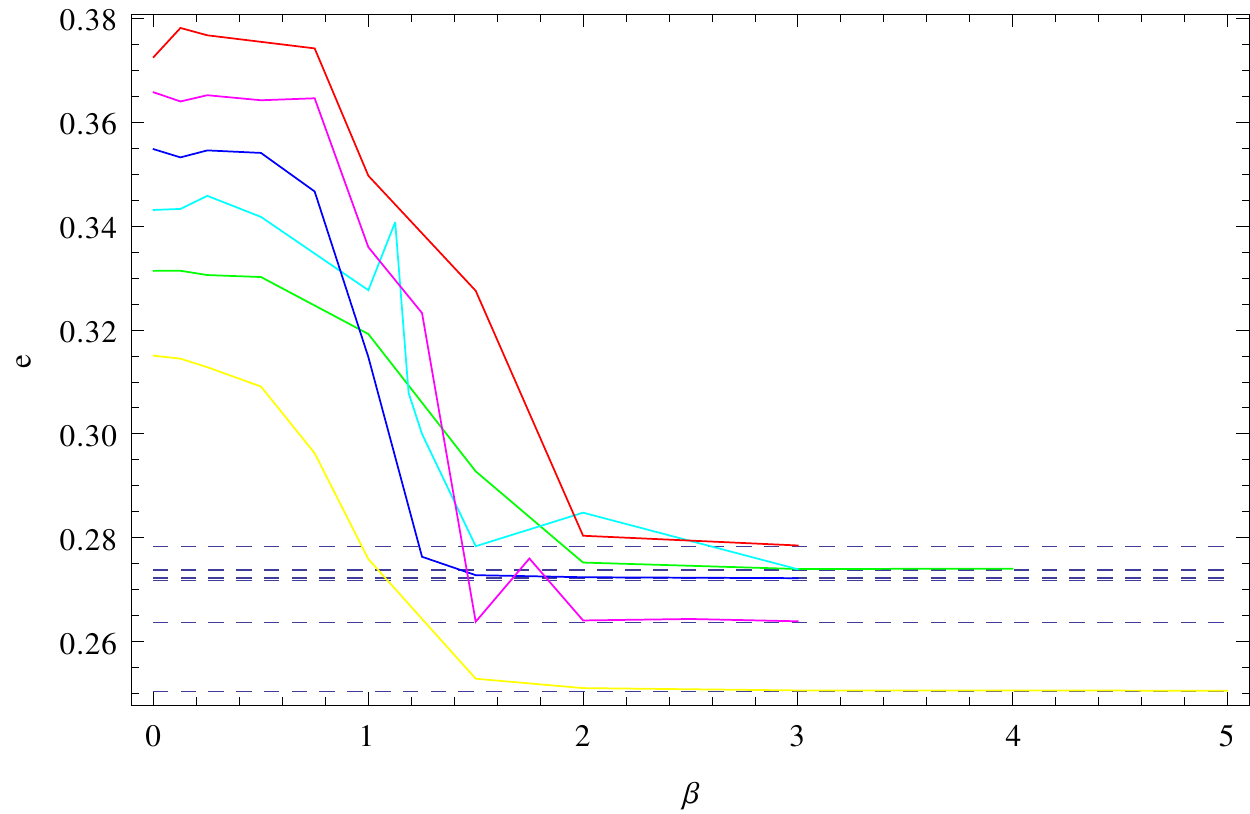}
\caption{Average energy $\langle e\rangle=-\frac{1}{d}\langle\log\varphi\rangle$ of a biased random walk as a function of temperature for dimensions $d=12-17$.}
\label{fig:mc12to17ebbeta}
\end{center}
\end{figure}
Such behavior suggests a phase transition as a function of $\beta$, as $d\to\infty$: as the temperature is lowered one leaves a \emph{universal} phase dominated by typical perfect lattices and enters a phase where lattices with low energies dominate the biased random walks. To test this assumption we define $\beta_c(d)$ as a solution to $E_d(\beta_c)) = E_c(d) = (E_d(0)+E_d(\infty))/2$. As usual $E_d(\infty)$ should read as $E_d(\beta_1)$ for some sufficiently large $\beta_1$. The crossover width is defined as $\beta_<(d)-\beta_>(d)$ where
\begin{gather*}
\Delta_d = \dfrac{E_d(0) - E_d(\infty)}{2}\\
E_<(d) = E_d(\beta_<) = E_d(\infty) + \dfrac{3}{4}\Delta_d = \dfrac{3}{4}E_d(0) - \dfrac{1}{4}E_d(\infty)\\
E_>(d) = E_d(\beta_>) = E_d(\infty) + \dfrac{1}{4}\Delta_d = \dfrac{1}{4}E_d(0) - \dfrac{3}{4}E_d(\infty)
\end{gather*}
The choice of factors $1/4$ and $3/4$ is not important and they can be replaced by other number.
\comm{\begin{figure}[htbp]
\begin{center}
\includegraphics[width=0.9\columnwidth]{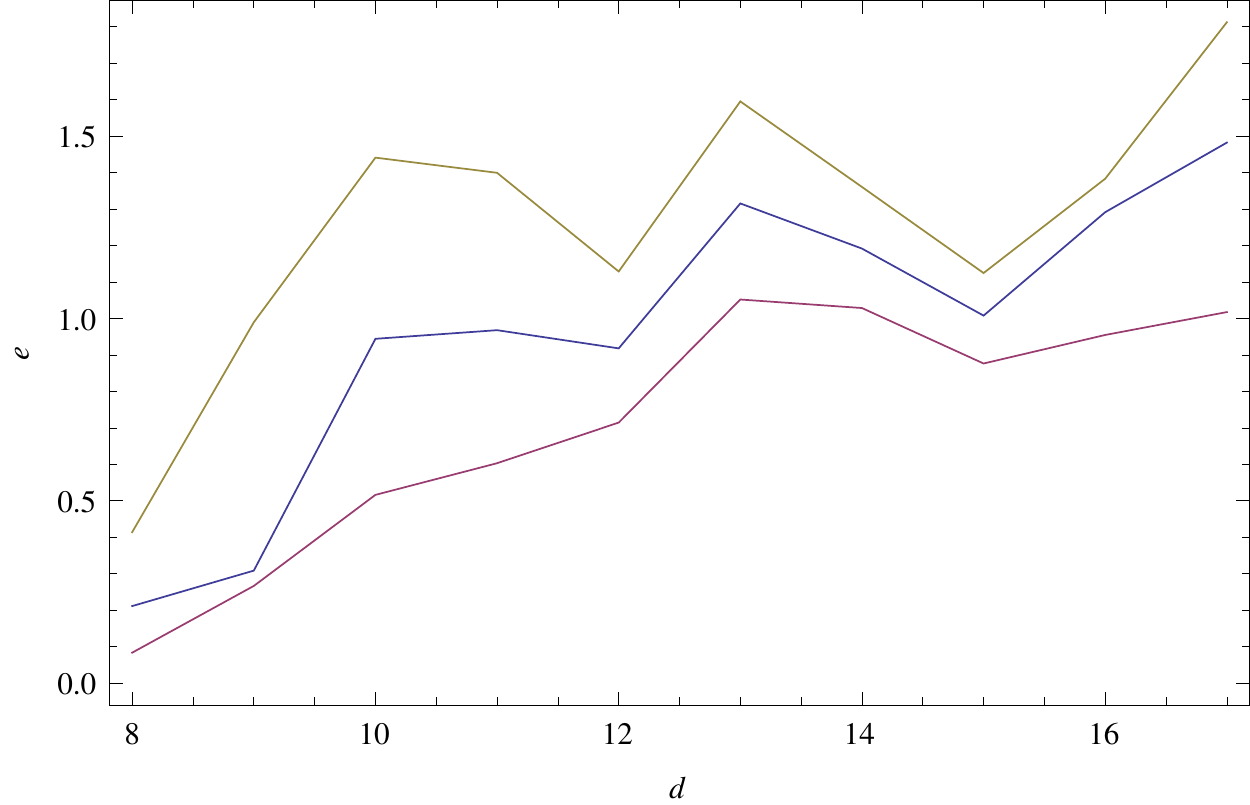}
\caption{$E_c$, $E_>$ and $E_<$ as functions of dimension ($d=8-17$).}
\label{fig:mc8to18ebsbcbl}
\end{center}
\end{figure}}
If there is indeed a phase transition then $W=(\beta_>-\beta_<)/\beta_c$ should converge to a constant value as $d\to\infty$. Fig.~\ref{fig:mc8to18ebscales} shows dependence of $W$ on dimension. One observes indeed a tendency to convergence to a constant value of $\Ord{1}$ (although with noticeable oscillations around it). We attribute the increase for $d>17$ to the glassy nature of the energy landscape of perfect lattices: these are exactly the dimension where the simple Monte-Carlo approach starts experiencing problems finding the best packer. The $d=18$ is intermediate between $d<18$ and $d=19$.

The situation seems to change qualitatively in $d=19$: for mildly low temperature one has to increase drastically  running time (as compared to $d<18$) in order to reach the best known packings. For very low temperatures, $\beta\sim3-5$ for $d=18,19$, the Monte-Carlo routine gets stuck around some relatively dense lattices and is never able to recover the densest lattice, or even approach it within the accuracy achieved in smaller dimensions. Typical energies reached by Monte-Carlo are of order $e\sim0.35-0.36$ for $\beta\lesssim5$. This is to be compared to the ground state $e=0.29$ corresponding to lattice $\Lambda_{19}$. It is then crucial to study higher dimensions in order to understand whether this behavior is a peculiarity of $d=19$ or it is a generic trend establishing in high dimensions. However we are unfortunately currently unable to investigate dimensions higher than 19 but we hope to be able to do so in the future.

\section{Diameter of the Voronoi graph}

An interesting question is the number of perfect forms as a function of dimension $d$. The exact numbers for $d<9$ and the estimate in $d=9$ suggest very steep, perhaps superexponential law which would make the full enumeration impossible beyond $d\sim11$. We conjecture that the number of perfect lattices should grow as $\mathcal{N}_d\sim\exp(A\,d^2)$ for an appropriate constant $A$ for large $d$. This conjecture is natural in the framework of statistical mechanics as the number of degrees of freedom is $\Ord{d^2}$ and so should be the ``entropy" of the system.

Looking at the distribution of the coefficients we can moreover conjecture that the Voronoi graph is a scale-free random graph, at least for a range of connectivities and for large $d$. For scale-free networks an estimate of number of vertices as a function of connectivities $c$ of the vertices of the graph is~\cite{bollobas2004diameter}
\begin{equation}
\label{eq:graphsized}
\dfrac{\log\mathcal{N}_d}{\langle\log c\rangle}\simeq\text{Diam}(\mathcal{G}_d).
\end{equation}
Here $\text{Diam}(\mathcal{G}_d)$ is diameter of the graph: the longest among the shortest paths between any pair vertices. 

We have estimated the diameter of the Voronoi graph $\mathcal{G}_d$ using the information on the graph provided by the random walk. This contains partial information and serves just as an order of magnitude consideration so we must consider the dependence on the size of the sample. This computation becomes increasingly harder with growing $d$ and we have restricted the study to $d\leq 11$. 

\begin{figure}[htbp]
\begin{center}
\includegraphics[width=0.85\columnwidth]{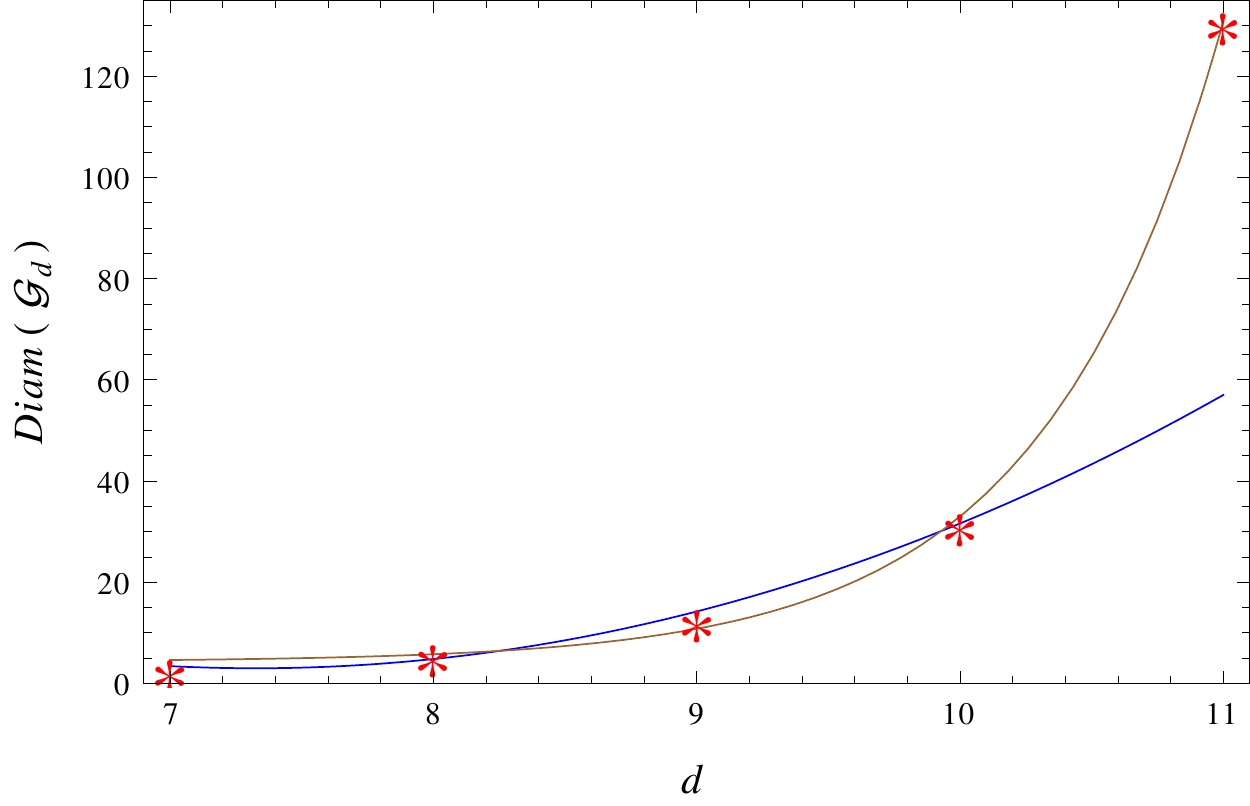}
\caption{\emph{Red crosses}: estimate of diameter of the Voronoi graph as function of dimension. \emph{Blue} and \emph{brown} curves are quadratic and exponential fits respectively provided here as guides for the eye.}
\label{fig:diamd}
\end{center}
\end{figure}

If the distribution of the connectivity is indeed scale free with fixed exponent $2.6$, we find that
\begin{equation}
\mean{\log c}=\frac{1}{2.6-1}=0.62,
\end{equation}
We find a reasonable agreement with numerical estimates of $\langle\log c\rangle$: $1.274$,$0.954$,$0.771$,$0.7$ for $d=8,9,10,11$ respectively. The excess of values of $\langle\log c\rangle_d$ with respect to conjectured value $0.62$ is due to the fact that we sample many well connected, dense lattices while not visiting many lattices with low connectivity. Therefore the logarithm of the size of the graph and the diameter should be proportional as
\begin{equation}
\log{\mathcal{N}_d}\simeq 0.62\ \text{Diam}(\mathcal{G}_d).
\end{equation}
We can then test if our hypotheses on the connectivity, the number of forms and size of the graph fit well together. We find graph diameters $3$, $6$, $13$, $32$ and $131$ for $d=7,8,9,10,11$ respectively. Remark that the exact diameter is $3$ and $4$ in $d=7$ and $8$ respectively. The growth is clearly faster than linear as shown on Fig~\ref{fig:diamd} and is consistent with the hypothesis of scale-free Voronoi graph. Quadratic fit for $\text{Diam}(\mathcal{G}_d)$ based on data for $d=7-10$ reads as:
\begin{gather*}
\text{Diam}(\mathcal{G}_d) = 217.6 -58.6\,d + 4\,d^2
\end{gather*}
However with the actual data we cannot find the precise scaling. Although exponential fit looks more accurate than quadratic on Fig~\ref{fig:diamd} we know that there are many forms in $d=11$  which were not visited by a random walk. Their addition to the graph would reduce the diameter and perhaps smear the seemingly exponential growth. More data are required to resolve this issue and we leave the resolution of this problem for future work.

\section{Trying to uniformize the choice of neighbor}

As we have already mentioned above, the randomization of Voronoi's algorithm is not unique: different cost functions~\eqref{eq:randomLP} produce slightly different results. We have considered a number of functions, targeting uniformization, i.e.\ trying to make sampling of rays/neighbors more uniform, more like it is for full enumeration. In all cases we observed a bias towards denser forms with higher kissing numbers, which we try to reduce. In particular we constructed a ``uniformized" cost function as shown on Fig.~\ref{alg:urrg} (recall that we have a $n$-dimensional polyhedron, $n = d(d+1)/2$ here, defined by a set of inequalities, the number of inequalities $\mathcal{N}\geq n$).
\begin{figure}
\begin{algorithmic}
\Require{Voronoi domain $\cV(Q)$}
\State Pick an inequality at random
\State Saturate the inequality, i.e. replace it with equality
\State Make a random Gaussian cost function $f$ as before
\State Solve linear program to get an extreme ray
\Ensure{Random extreme ray $R$}
\end{algorithmic}
\caption{Algorithm for uniformized random extreme ray generation.}
\label{alg:urrg}
\end{figure}
 This construction is inspired by the remark that purely random cost function generates rays weighted with areas of facets adjacent to that ray, and it also favors forms that have higher connectivity, i.e.\ number of neighbors. This is an advantage if one is interested in denser forms. However if one is studying properties of the Voronoi graph it might be preferable to make the outcome of neighbor generation more uniform.

\begin{figure}[htbp]
\begin{center}
\includegraphics[width=0.9\columnwidth]{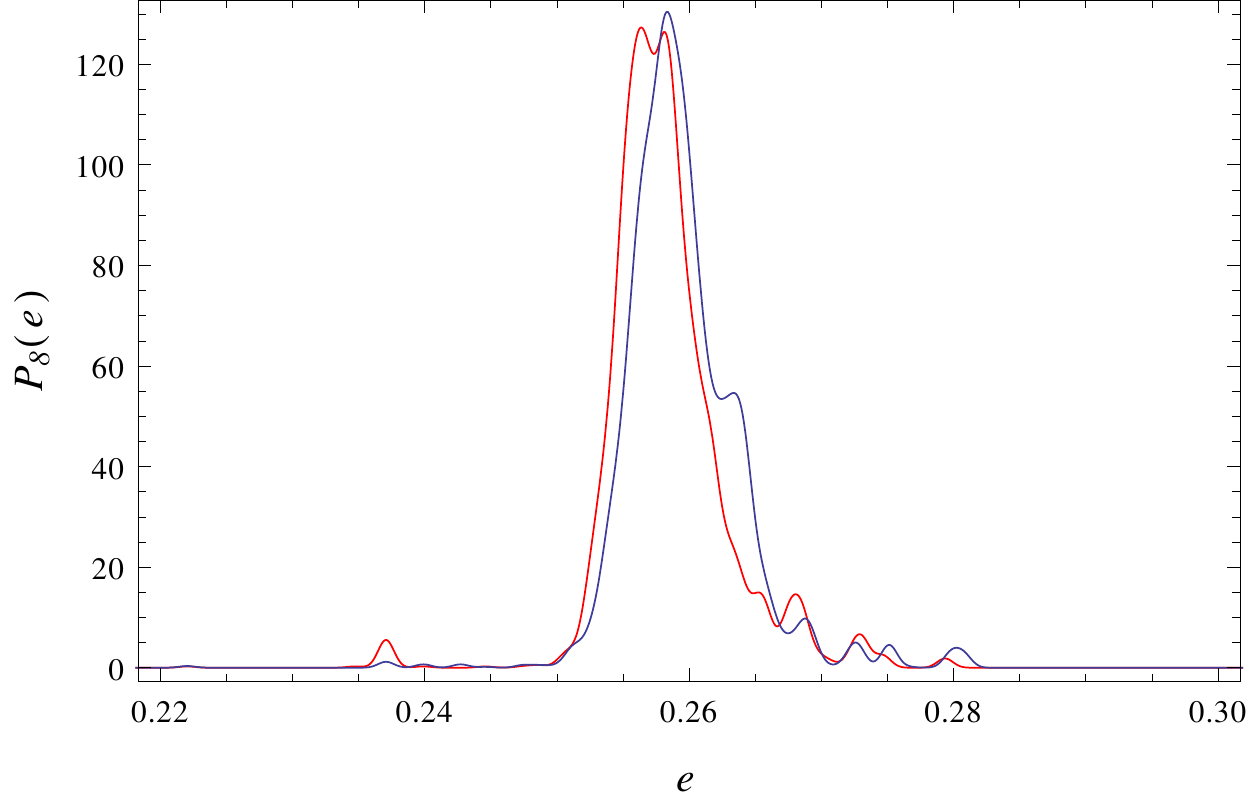}
\includegraphics[width=0.9\columnwidth]{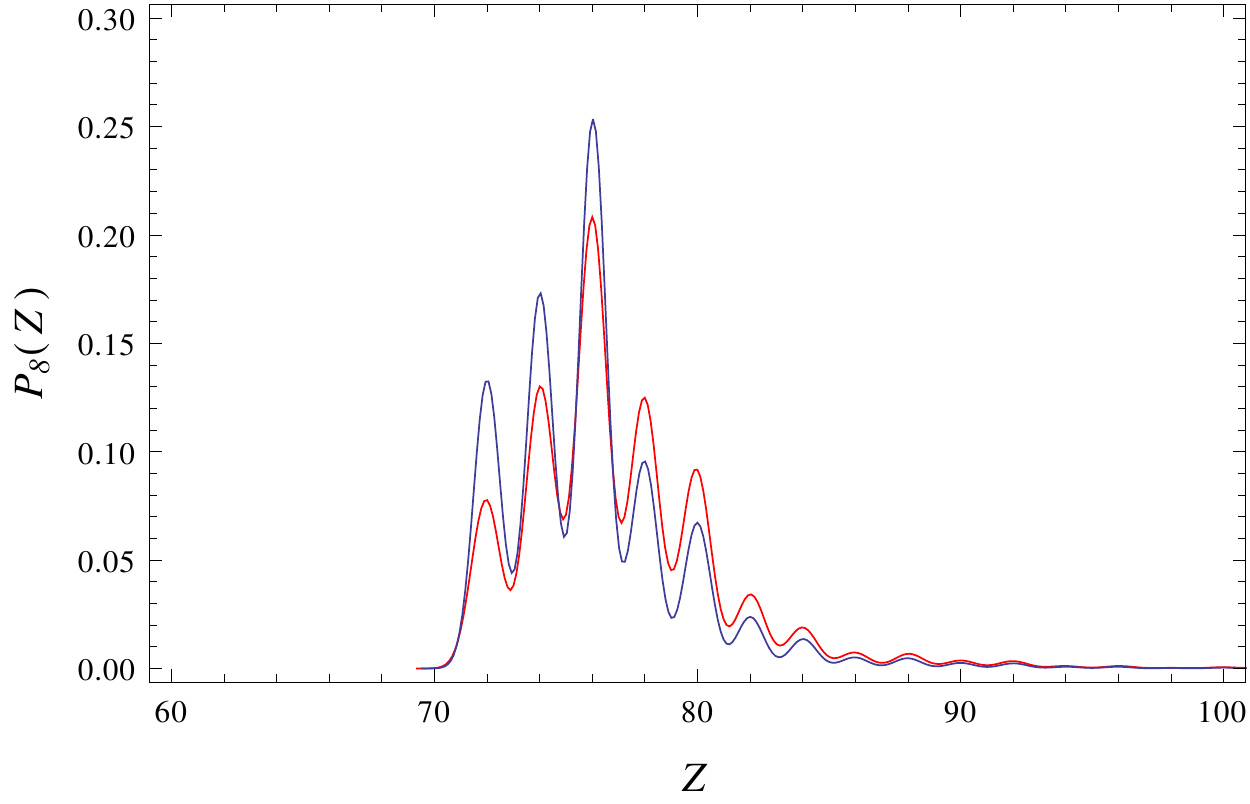}
\caption{\emph{Top.} Distribution of energies \emph{Bottom.} Distributions of kissing numbers. Blue and red curves are generated by random walks in $d=8$ with \emph{random} and \emph{uniformized} cost functions.}
\label{fig:e8-pdf-m1m5}
\end{center}
\end{figure}

The above construction tries to give facets a more uniform weights. Comparison of numerical results for random and uniform cost functions are presented on Fig.~\ref{fig:e8-pdf-m1m5} which shows distributions of kissing number and energy in $d=8$. There's no significant difference of distributions between the \emph{random} and \emph{unformised} cost function.
However the uniformized cost function is advantageous over the random function if one is interested in the properties of the Voronoi graph: typically it yields more non-isometric forms than the pure random function for equal number of runs. We have performed this comparison for $d=8-12$ and results are summarized in the table below (where Fraction column is a ratio $\mathcal{N}_u/\mathcal{N}_r$ of number of forms found $\mathcal{N}_r$ and $\mathcal{N}_u$ with \emph{random} and \emph{uniformized} cost functions respectively):
\begin{center}
\begin{tabular}{ | c | c | c | c | c | }
\hline
Dim. & Steps & Random & Uniformized & Fraction\\
\hline
$8$ & $2\,10^6$ & $1793$ & $2955$ & $1.648$ \\
$8$ & $4\,10^6$ & $2529$ & $3963$ & $1.567$ \\
\hline
$10$ & $10^6$ & $331065$ & $434317$ & $1.312$ \\
\hline
$11$ & $10^6$ & $744282$ & $825695$ & $1.109$ \\
\hline
\end{tabular}
\end{center}
The difference between the two cost functions is decreasing rapidly as dimensionality is increased. We believe that these strategies are better suited for lower dimensions $d\lesssim12$ where isometry is important.

\section{Extension to periodic sets}

Before concluding let us describe a possible extension of our approach to lattices with many particles per unit cell which we refer to as periodic sets throughout this section. Such an extension is possible but has a number of limitations which make the problem more difficult than the Bravais lattice version.

The generalization of the Voronoi algorithm to periodic sets was introduced by Sch\"urmann~\cite{schurmann2009computational}. An $m$-periodic set is defined by a quadratic form which describes how a unit cell is translated in space and a set of $m$ real vectors (\emph{translational part}) that defines the positions of $m$ particles inside the cell. It is then possible to extend the Voronoi theory presented in Sec.~\ref{sec:voronoi} and introduce $m$-perfect and $m$-eutactic lattices; \emph{$m$-extreme} lattices are defined as local maxima of packing fraction of $m$-periodic sets, just like in the Bravais case. There is as well an analogue of the Ryshkov polyhedron.

It is at this point that a crucial difference appears which makes the problem more complicated than the lattice one. In general not all extreme lattices are $m$-perfect and $m$-eutactic: there exist lattices which are extreme, but not perfect. An example is provided by \emph{fluid diamond packings}~\cite{conway1999sphere} where a fraction of spheres can be moved around freely without canghing the packing fraction. Furthermore the Voronoi graph no longer exists: the method only provides a local direction in which packing fraction is increasing. Potentially this allows to design an algorithm that starts with a periodic set and end up at an $m$-perfect lattice~\cite{schurmann2009computational}. On the other hand, the extension to many particles in a unit cell highlights the importance of perfect, strongly eutactic lattices since one can prove that they are \emph{extreme}~\cite{schurmann2010perfect}, that is they are extreme among lattices with any number of particles per unit cell.

These limitations are lifted if one fixes translational part and replaces \emph{real} vectors in the definition of a peridic set by their \emph{rational} approximations~\cite{schurmann2009computational}. Under this assumption, all the features of the Voronoi theory are recovered. Yet the complexity is increasing too: the computation of the shortest vectors of such periodic set is more involved.

\section{Conclusions and further directions}

We have suggested a new approach to the lattice sphere packing problem based on randomization of the Voronoi algorithm. Previous works used complete enumeration that becomes computationally unfeasible beyond $d\sim10-11$ (see however~\cite{bremner2009polyhedral,sikiric2009complexity,dutour2010contact,rehn2011c++}). We have developed an implementation of our algorithm that allowed us to study dimensions from $8$ to $19$ and we foresee its application for studying perfect lattices up to $d=40$ at least (beyond that, technical problems with the implementation of the algorithm become conceptual problems). 

We have studied statistical properties of the sets of perfect lattices generated by our algorithm, both typical and extreme values focusing on two quantities: energy, which we define as proportional to the logarithm of the packing fraction, and the kissing number. For all dimensions except $d=19$ we were able to retrieve the best known packings starting from $A_d$ or $D_d$ lattices either using simple random walk for $d\leq12$ or biasing the random walk with temperature for $d>12$. In $d=19$ we had to restart the walk many times in order to hit the best packer: random walk was always  getting stuck in some higher-energy lattice, a phenomenon which is reminiscent of a glassy free energy landscape. The change of the average energy with temperature suggests the existence of a sharp phase transition as $d\to\infty$, although we cannot argument on this topic more, due to the large dimension-dependent fluctuations as the energy is lowered. We do not exclude we will be able to say more on this topic in future work. 

We also found that the typical values tend to have much smoother behavior what allowed us to propose two possible scenarios for the large $d$ behavior of the packing fraction of the typical perfect lattices: in one case we obtain en exponential decay of the packing fraction whose leading order improves upon Minkowsky's bound
\begin{equation}
\phi\sim 2^{-(0.84\pm 0.06) d},
\end{equation}
while in the second case we have a faster, factorial-like decay
\begin{equation}
\phi\sim d^{-(0.06\pm 0.04)d}
\end{equation}
however with an unnaturally small exponent. The resolution of this conundrum would need investigation of lattices in dimensions 40 and higher.

Higher dimensions  are also accessible and will require mostly technical rather than conceptual modifications in the code, at least for $d\leq 40$. Getting beyond $d=24$ is quite important since in dimensions below $24$ are dominated by the Leech lattice $\Lambda_{24}$ and all the densest lattices in these cases are cross sections of $\Lambda_{24}$.

Other possible applications of our work include a test of the ``decorrelation principle" in \cite{torquato2006new}, by studying the two-particles correlation functions of typical perfect lattices, and a systematic study of the \emph{perfect and eutactic} lattices which are the true local minima of the energy for the purpose of unveiling a glassy structure of the energy landscape. Checking for eutaxy is quite straightforward, after a set of perfect lattices has been generated, but we found that this requires a much larger statistics than that used in our paper since the rejection rate is quite large: as dimension of space is increased the fraction of (at least) eutactic lattices discovered by a plain random walk drops rapidly as illustrated in Table~\ref{tab:eut8to19}. If one biases the walk with temperature the numbers increase, but they are still low and we have not tested whether the increase is due to different lattices or isometric copies of few lattices. Therefore we leave this for future work.
\begin{table}
\begin{tabular}{ | c | c | }
\hline
Dimension & Fraction of eutactic lattices\\
\hline
8 & 0.997 \\
\hline
9 & 0.830 \\
\hline
10 & 0.738 \\
\hline
11 & 0.479 \\
\hline
12 & 0.134 \\
\hline
13 & 5.11e-03 \\
\hline
14 & 3.00e-04 \\
\hline
15 & 1.30e-04 \\
\hline
16 & 8.00e-05 \\
\hline
17 & 6.00e-05 \\
\hline
18 & 1.25e-05 \\
\hline
19 & 2.00e-05 \\
\hline
\end{tabular}
\caption{Fraction of eutactic and strongly eutactic discovered by random walk for $d=7-19$.}
\label{tab:eut8to19}
\end{table}

Finally, randomization procedure we have introduced could also be applied to other optimization problems like lattice covering problem~\cite{schurmann2009computational}, where one searches for the most economical way of covering a space with spheres of equal size. Another possible activity along the same direction is to adapt our randomization procedure to the algorithm generating all eutactic lattices in a given dimension~\cite{batut2001classification}.

As we have indicated, finding extreme rays of the Voronoi domain $\cV$ is a particular case of a general \emph{polyhedral representation conversion problem}~\cite{avis2009polyhedral}. This is an important problem in \emph{combinatorial optimization} and \emph{computational geometry}. Although efficient algorithms exist for certain classes of polyhedra, its complexity in general is unknown~\cite{avis2009polyhedral,bremner2009polyhedral} but all existing algorithms, that perform the full conversion, are exponential in dimension of a polyhedron~\cite{avis2009polyhedral}. In this wider context our randomization approach offers a possible workaround for optimization problems which require solution of the representation conversion problem in order to find an optimum.

\section{Acknowledgements}

We wish to thank A.Sch\"urmann and G.Nebe for providing code for isometry testing. We are also indebted to S.Torquato, A. Kumar and H. Cohn for many stimulating discussions. We would like to thank the developers of PARI/GP~\cite{PARI2} libraries for their quick response in fixing bugs. AS would like to thank the Center for Theoretical Physics at MIT where part of this work was completed.

\section*{Appendix A. Some technical details}

The two main techincal ingredients of the Voronoi algorithm are generation of random extreme ray $R$ of the Voronoi domain $\cV(Q)$ and finding a neighbour $Q^\prime$ of a given lattice $Q$ provided an extreme ray $R$.

Computing a random extreme ray has the same complexity as generating the Voronoi domain $\cV(Q)$ and solving a linear program. We need to know shortest vectors of $Q$ in order to build $\cV(Q)$. Computing shortest vectors of a lattice is exponentially hard problem in $d$. However decent algorthims exist allowing computation to caried out in reasonable time at least up to $d\sim40$~\cite{fincke1985improved,cohen1993course}. The other source of complexity is the size of linear program which is defined by kissing number of $Q$ (and hence scales exponentially in $d$ for dense packings) and is limited by ability of linear program (LP) solvers to cope with huge linear programs: size of LP becomes of order $10^{10}$ for the densest known lattices in $d\gtrsim 40$. Based on this observations we expect our method to work up to $d\sim 40$, at least in theory. It is also worth pointing that it is straightforward to check if a given ray $R$ is extreme~\cite{avis2009polyhedral}.

Finding a neighbour $Q^\prime = Q + \alpha\,R$ with $\alpha\in\mathbb{Q}$ proved to be a harder problem computationally and it is this part of the problem that put limited our data by $d<20$. The value of $\alpha$ is rational~\cite{schurmann2009computational,martinet2003perfect}, so that we can always choose $Q^\prime$ to be integral and all perfect lattices then have integral representation. We use modified binary search algorithm as defined by Sch\"urmann~\cite{schurmann2009computational} to compute neighbours of a lattice ($S_{>0}^d$ is set of all lattices) presented on Fig.~\ref{alg:mbs}. 
\begin{figure}
\begin{algorithmic}
\Require{perfect form $Q$, extreme ray $R$}
\While{$Q+u\,R\not\in S_{>0}^d\text{ and }\lambda(Q+u\,R) = \lambda(Q)$}
\If{$Q+u\,R\not\in S_{>0}^d\text{ and }\lambda(Q+u\,R) = \lambda(Q)$}
\State $u\leftarrow (l+u)/2$
\Else
\State $(l,u)\leftarrow(u,2\,u)$
\EndIf
\EndWhile
\While{$\text{Min}(Q + l\,R)\subset\text{Min}(Q)$}
\State $g\leftarrow(u+l)/2$
\If{$\lambda(Q + g\,R)\geq\lambda(Q)$}
\State $l\leftarrow g$
\Else
\State $u\leftarrow\min\{(\lambda(Q) - Q[v])/R[v] | v\in\text{Min}(Q+g\,R), R[v] < 0\}\cup\{g\}$
\EndIf
\EndWhile
\Ensure{$\alpha\leftarrow l$}
\end{algorithmic}
\caption{Modified binary search for neighbour $Q^\prime$ of a lattice $Q$ given an extreme ray $R$.}
\label{alg:mbs}
\end{figure}
 The idea behind this construction is very simple: the neighbour of $Q$ is $Q^\prime = Q + \alpha\,R$ with the smallest positive rational $\alpha$ such that $\lambda(Q) = \lambda(Q + \alpha\,R)$ and $\text{Min}(Q + \alpha\,R)\not\subseteq\text{Min}(Q)$~\footnote{Note that self-loops are allowed, i.e. $\alpha\neq 0$ and $\text{Min}(Q + \alpha\,R)=\text{Min}(Q)$.} In the first part above upper and lower boundaries for $\alpha$ are defined. The second part is a modified binary search for value of $\alpha$. The modification - an extra conditional in the assignment of $u$ - is necessary to make the algorithm converge in finite number of steps to an exact rational value of $\alpha$.

\section*{Appendix B. Random walks and isometry check}

We have used two different approaches to perform checks for isometry of lattices. In the first approach we split the data generation in two steps
\begin{itemize}
\item Generate a random walk in space of lattices with no check for isometry.
\item Run isometry test on the trajectory of the random walk and generate an approximate Voronoi's graph.
\end{itemize}
After the first step one obtains a full trajectory of a random walk as list of lattices. The second step generates the graph by eliminating isometric copies of lattices by glueing together isometric elements of the list. This induces a relation of neighbourhood in the list and transforms the list into a graph. Second possibility is to perform isometry check and graph construction on the fly ($P\sim Q$ denotes isometric equivalence, $V$ and $E$ are sets of vertices and edges of the graph $G$ respectively) as shown on Fig.~\ref{alg:svg}.
\begin{figure}
\begin{algorithmic}
\Require{perfect $Q$, graph $G=(V=\emptyset,E=\emptyset)$}
\Loop
\State Random extreme ray $R\leftarrow Q$
\State Neighbour $Q^\prime\leftarrow Q + \alpha\,R$
\For{$P\in G$}
\If{$P\sim Q^\prime$}
\State $E\leftarrow E\cup(Q,P)$
\Else
\State $V\leftarrow V\cup Q^\prime$
\State $E\leftarrow E\cup (Q^\prime,Q)$
\EndIf
\EndFor
\EndLoop
\Ensure{Voronoi graph $G$}
\end{algorithmic}
\caption{Algorithm that constructs an approximation to the Voronoi graph}
\label{alg:svg}
\end{figure}
 Algorithm terminates after a predefined number of steps has been done.

An algorithm to check whether two lattices are isometric was developed by W. Plesken and B. Souvignier in Ref.~\cite{plesken1997computing}. We adapted the original code of B. Souvignier to perform isometry testing.

\section*{Appendix C. Na\"ive random walk}

It is worth discussing performance of a straightforward approach one might be tempted to follow. The Voronoi construction is elaborate and requires computational effort. \emph{A priori} one might wonder if a simple \emph{lattice random walk/Monte-Carlo} is preferrable (maybe in higher dimensions) ? The algorithm shown on Fig.~\ref{alg:nrw} is extremely simple: one hopes to approach the best packer by small steps if the random walk is sufficiently biased towards denser lattices. When generating a move one has the option of eigther producing a new lattice $A^\prime$ which might or might not be an isometric copy of $A$. Acceptance probability $p$ could be $1$ (random walk) or for example, Metropolis rule (\`a la Monte-Carlo).

\begin{figure}
\bigskip
\begin{algorithmic}
\Require{Lattice $A$}
\Loop
\State $(*)\quad A^\prime\leftarrow A$
\State Accept $A^\prime$ with some probability $p$
\State Goto $(*)$
\EndLoop
\Ensure{Dense lattice $A$}
\end{algorithmic}
\caption{Na\"ive random walk}
\label{alg:nrw}
\end{figure}

Unbiased random walk (infinite temperature in our language) with moves that generate non-isometric lattices $A^\prime$ gives an average packing fraction which is equal to Minkovsky's bound~\cite{parisi2008most,rogers1958packing,rogers1955mean,rogers1956number,rogers1964packing}. This is a rather strong result since Minkovsky's bound is non-constructive and constructing a lattice in a given dimension satisfying the bound is yet an open problem. However it is very hard to implement that type of updates in practice~\cite{parisi2008most,siegel1945mean,rogers1955mean,rogers1956number,rogers1958packing,rogers1964packing} and one has to rely on various approximations. In case when one allows for any $A^\prime$ the perfomance of the algorithm is extremely poor: with the simple Gaussian measure for lattices~\cite{parisi2008most} $\mathcal{P}(A)\sim\exp(-\gamma\,\text{Tr}\,AA^t)$ we were able to recover the best packers in $d=2,3$, although already in $3$ dimensions we had to go to very low termperatures. The performance of the algorithm quickly deteriorates with dimensions, and by $d=10$ it is completely useless. The above mentioned variant of the algorithm where one samples only among the non-isometric lattices has similar performance when approximantions are used. Finally it's worth mentioning that the lattices generated by such Markov chains are never perfect and are typically far from being such.

These negative results provide an extra motivation for studying perfect lattices and the Voronoi construction where much better performance is achieved.

\bibliography{Voronoi}{}

\end{document}